\documentclass{article}

\usepackage{a4wide}
\usepackage{natbib}

\usepackage{timet}
\usepackage{amsmath}
\usepackage{amsfonts}
\usepackage{amssymb}
\usepackage{graphicx}
\usepackage{subfigure}   
\usepackage{hyperref}  

\newcommand{\vect}[1]{{ \boldsymbol{#1} }}
\newcommand{\unit}[1]{\hat{\boldsymbol{#1}}}

\title
{Precessing spherical shells: flows, dissipation, dynamo and the lunar core}

\author 
{D. C\'ebron$^1$\thanks{david.cebron@univ-grenoble-alpes.fr}, R. Laguerre$^2$, J. Noir$^3$, N. Schaeffer$^1$\thanks{nathanael.schaeffer@univ-grenoble-alpes.fr} \\
  \small $^1$ Universit\'e Grenoble Alpes, CNRS, ISTerre, Grenoble, France\\
  \small $^2$ Royal Observatory of Belgium, Brussels, Belgium\\
  \small $^3$ Institute of Geophysics, ETH Zurich, Zurich, Switzerland
  }
\date{\today}

\begin{document}

\maketitle

\begin{abstract}
\normalsize
Precession of planets or moons affects internal liquid layers by driving flows, instabilities and possibly dynamos.
The energy dissipated by these phenomena can influence orbital parameters such as the planet's spin rate.
However, there is no systematic study of these flows in the spherical shell geometry relevant for planets, and the lack of scaling law prevents convincing extrapolation to celestial bodies.

We have run more than 900 simulations of fluid spherical shells affected by precession, to systematically study basic flows, instabilities, turbulence, and magnetic field generation.
We observe no significant effects of the inner core on the onset of the instabilities.
We obtain an analytical estimate of the viscous dissipation, mostly due to boundary layer friction in our simulations.
We propose theoretical onsets for hydrodynamic instabilities, and document the intensity of turbulent fluctuations.

We extend previous precession dynamo studies towards lower viscosities, at the limits of today's computers.
In the low viscosity regime, precession dynamos rely on the presence of large-scale vortices, and the surface magnetic fields are dominated by small scales.
Interestingly, intermittent and self-killing dynamos are observed.
Our results suggest that large-scale planetary magnetic fields are unlikely to be produced by a precession-driven dynamo in a spherical core.
But this question remains open as planetary cores are not exactly spherical, and thus the coupling between the fluid and the boundary does not vanish in the relevant limit of small viscosity.
Moreover, the fully turbulent dissipation regime has not yet been reached in simulations.

Our results suggest that the melted lunar core has been in a turbulent state throughout its history.
Furthermore, in the view of recent experimental results, we propose updated formulas predicting the fluid mean rotation vector and the associated dissipation in both the laminar and the turbulent regimes.
\end{abstract}


\section{Introduction}

The origin of the magnetic fields of planets and stars is attributed to the dynamo mechanism. It is commonly thought that most of the dynamos are powered by compositional and thermal convection in the liquid part of these objects. Nevertheless, this scenario is sometimes difficult to apply.
This is for instance the case for the early Moon, for which the intensity of the magnetic field generated by convection might not be sufficient \citep{stegman2003early} or the Earth, where recent estimates of thermal and electrical conductivity of liquid iron imply that convection would be far less efficient than previously thought \citep{pozzo2012thermal}. Mechanical forcings constitute then  alternative ways to sustain dynamo action \citep{le2015flows}, as shown numerically for libration \citep{wu2013dynamo}, tides \citep{cebron2014tidally,vidal2017tidala} or precession.
The present study focuses on precession, which has already been demonstrated numerically to be able to grow a magnetic field in spherical shells \citep{tilgner2005precession,tilgner2007kinematic}, full spheres \citep{lin2016precession}, cylinders \citep{nore2011nonlinear,cappanera2016,giesecke2018nonlinear} and cubes \citep{goepfert2016dynamos,goepfert2018mechanisms}.
Hence, the possibility of a precession driven dynamo in the liquid core of the Earth \citep{kerswell1996upper} or the Moon \citep{dwyer2011long} cannot be excluded.
However, current numerical simulations operate at viscosities many orders of magnitude higher than natural dynamos.
The present work aims at shedding some light on the consequences of precession in spheres, including dissipation and magnetic field generation.
To this end, we make extensive use of numerical simulations pushing down the viscosity to the limits of current supercomputers.

A rotating solid object is said to precess when its rotation axis itself rotates about a secondary axis that is fixed in an inertial frame of reference.
The first theoretical studies of precession considered an inviscid fluid \citep{hough1895oscillations,sloudsky1895rotation,poincare1910precession}. Assuming a uniform vorticity, they obtained a solution for the spheroid, called Poincar\'e flow, given by the sum of a solid body rotation and a potential flow. However, the Poincar\'e solution is modified by the existence of boundary layers, and some strong internal shear layers are also created in the bulk of the
flow \citep{stewartson1963motion}. In 1968, Busse took into account these viscous effects as a correction to the inviscid flow in a spheroid, by
considering carefully the Ekman layer and its critical regions
 \citep{busse1968steady,zhang2010fluid}. Based on these works, \cite{cebron2010tilt} and \cite{Noir2013} have proposed models for the flow forced in precessing triaxial ellipsoids.
Beyond this correction approach, the complete viscous solution (including the fine description of all viscous layers) has been obtained for the sphere in the two limit cases of a weak \citep{kida2011steady} and strong \citep{kida2018steady} precession rates. 

When the precession forcing is large enough compared to viscous effects, instabilities can occur in precessing spherical containers, destabilizing the Poincar\'e flow (e.g. \citet{hollerbach2013parity}). First, the Ekman layers can be destabilized \citep{lorenzani2001fluid2} through standard Ekman layer instabilities \citep{lingwood1997absolute,faller1991instability}. In this case, the instability remains localized near the boundaries. Second, the whole Poincar\'e flow can be destabilized, leading to a volume turbulence : this is the precessional instability \citep{malkus1968precession}. 
It has been argued by \cite{lorenzani2001fluid2}, and more recently  by \cite{lin2015shear}, that the conical shears spawned at the critical latitudes can couple non-linearly with pairs of inertial modes, leading to the Conical Shear Instability (CSI) of \cite{lin2015shear}.

In this work, we will study the influence of the presence of a solid inner core on the precessional instability.
We thus consider a spinning and precessing spherical shell filled with a conducting fluid.
Adopting the approach used for ellipsoids by \cite{Noir2013}, we obtain an explicit expression of the fluid rotation vector in the presence of an inner core.
We then derive hydrodynamical stability criteria for the CSI involving conical shear layers spawned by the outer and the inner spherical shell.
Finally, based on our hydrodynamic simulations, we investigate precession driven dynamos in different flow regimes.
The pioneering results obtained by \cite{tilgner2005precession} are extended to smaller viscosities, in the hope to reach an asymptotic regime relevant for planetary fluid layers.

Our paper is organized as follow. Section \ref{description_pb} presents the governing equations of the problem and a brief description of the numerical method used to solve the equations.
A reduced model for the base flow in  spherical shells is then derived and compared to numerical simulations in Section \ref{base_flow}, while transition to unstable flows is studied in Section \ref{hydro_instability}.
Precession driven dynamos are studied in Section \ref{dynamo}.
Finally, we apply our findings to the Moon (\S\ref{appliMoon}) and draw some conclusions (\S\ref{conclusion}).

\section{Description of the problem and mathematical background}
\label{description_pb}

\begin{figure}                
  \begin{center}
     \includegraphics[width=0.6 \linewidth]{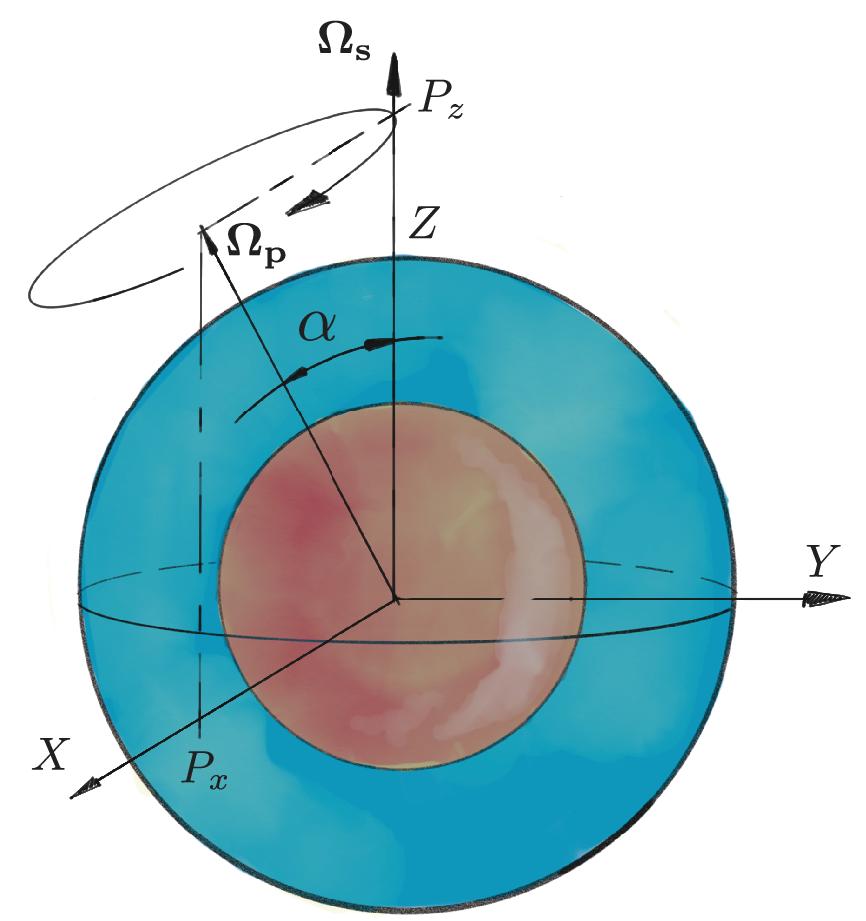}  
  \caption{Schematic description of the problem.}
  \label{fig:setup}                
  \end{center}
\end{figure}

We consider an incompressible Newtonian fluid of density $\rho$, kinematic viscosity $\nu$, electrical conductivity $\gamma$, and magnetic permeability $\mu$, enclosed in a spherical shell of outer radius $R$ and inner radius $R_i$. In the following, the outer boundary is also named CMB, for core-mantle boundary, whereas the inner one is also called ICB, for inner core boundary. When present, the ICB rotation vector is assumed to be the same than the CMB one (thus precessing with the mantle). The cavity rotates with an angular velocity $\vect{\Omega_s}=\Omega_s \unit{z}$ and is in precession at $\vect{\Omega_p}=\Omega_p \unit{k}$, with the unit vector $\unit{k}$ and $\unit{k}\cdot \unit{z}=\cos \alpha$. We define the frame of precession as the frame of reference precessing at $\Omega_p$, in which we construct a Cartesian coordinate ($\unit{x},\unit{y},\unit{z}$) system centered on the sphere, with $\unit{z}$ along $\vect{\Omega_s}$, $\unit{x}$ such that $\vect{\Omega_p}$ is in the plane xOz and $\unit{y}=\unit{z}\times\unit{x}$ (Fig. \ref{fig:setup}).  

\subsection{Mathematical formulation}

Defining $ \Omega_o = \Omega_s+\Omega_p$, we choose  $\Omega_o^{-1}$ as the unit of time such that it remains relevant in both limits of large $\Omega_s$ or large $\Omega_p$ \citep[see also][]{goepfert2016dynamos}.
We choose  $R$ and $R \Omega_o \sqrt{\mu \rho}$ as the respective units of length and magnetic field. In the frame of reference precessing at $\vect{\Omega}_p$, the dimensionless governing equations take the form:  

\begin{eqnarray}
\frac{\partial \vect{u}}{\partial t} +\vect{u}\cdot\nabla \vect{u} &=& -\nabla p+E\, \nabla^2 \vect{u}-2\frac{Po}{1+Po}\, \unit{k} \times\vect{u}+\bf(\nabla\times B)\times B, \label{NS} \\
\nabla \cdot \vect{u} &=& 0 \label{eq:divu}  \\
 {\partial{\bf B}\over\partial t}&=& \frac{E}{Pm}\,\nabla^2{\bf B} \label{eq:induction}
  + \nabla\times({\bf u\times B}), \\
    \nabla \cdot {\bf B} &=&0, \label{eq:divB}
\end{eqnarray}
where $p$ is the reduced pressure accounting for centrifugal forces (in our simulations, $p$ is eliminated by taking the curl of equation \ref{NS}). The four dimensionless parameters controlling the dynamics of the system are the Ekman, magnetic Prandtl, Poincar\'e numbers and aspect ratio, respectively defined by $E=\nu/(\Omega_o R^2)$, $Pm=\mu \gamma \nu$, $Po=\Omega_p/\Omega_s$, and $\eta=\frac{R_i}{R}$. 

Note that with our choice of time scale, the instantaneous rotation vector of the spherical container is $\vect{\tilde{\Omega}_s}=(1+Po)^{-1} \unit{z}$. Hence, our modified Ekman number is related to the more classical definition based on $\Omega_s$, $E'=\nu/(\Omega_s R^2) = E (1+Po)$.
Note also that, in this study, our convention is to consider positive $Po$ and $\alpha>90^{\circ}$ for retrograde precession, while others consider negative values of $Po$ with $\alpha \in [0,90^{\circ}]$ \citep[e.g.][]{dwyer2011long}.

We also use the spherical coordinate system $(r,\theta,\phi)$, $r$ being the radial distance, $\theta$ the colatitude, and $\phi$ the azimuthal angle (with Ox the axis of zero longitude).

\subsection{Numerical approach} \label{sec:num}

We impose no-slip boundary conditions at both boundaries, i.e. $\vect{u}= \vect{\tilde{\Omega}_s} \times \vect{r}$ in our precessing frame of reference. We consider insulating boundary conditions for the magnetic field at the CMB and an inner core electrical conductivity equal to that of the fluid.
The problem is solved using the XSHELLS code (freely available at \url{https://bitbucket.org/nschaeff/xshells}).
This high performance, parallel Navier-Stokes solver works in spherical coordinates using a toroidal-poloidal decomposition and a pseudo-spectral approach.
Note that with this approach, equations (\ref{eq:divu}) and (\ref{eq:divB}) are automatically satisfied.
The spherical harmonic transforms are performed using the efficient SHTns library \citep{schaeffer2013efficient} while the radial direction is discretized with second order finite differences.
XSHELLS has been benchmarked on convective dynamo problems with or without a solid inner-core \citep{marti2014,matsui2016}.
\begin{figure}                
  \begin{center}
   \includegraphics[width=0.7 \linewidth]{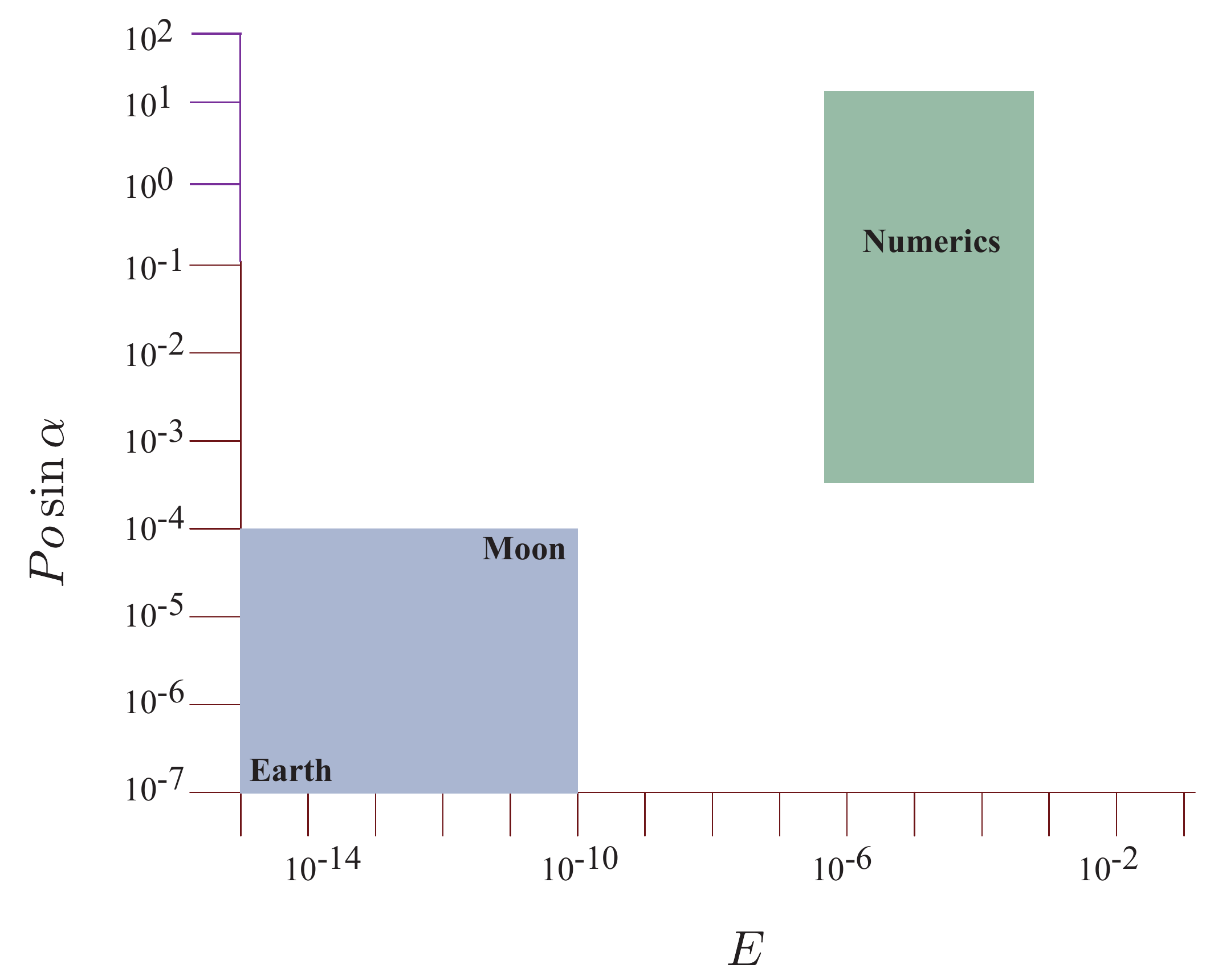}
  \caption{Parameter space covered by numerical studies and for typical planetary cores (opposite corners show here the current Earth's and Moon's liquid cores).}
  \label{fig:parameterRange}                
  \end{center}
\end{figure}   
In the following, the so-called hydrodynamic simulations do not take into account the magnetic field (${\bf B}$), whereas our so-called magnetohydrodynamic (MHD) simulations solve the full system (\ref{NS})-(\ref{eq:divB}). The range of hydrodynamic parameters investigated in this study are summarized in Fig. \ref{fig:parameterRange} and compared to the typical values expected for planetary cores. In addition we explore the range $0.2 < Pm < 20$ in our MHD simulations.

We checked on a few cases that computing in the mantle frame (i.e. the frame co-rotating with the solid shell) gives the same result as in the precession frame.
Note that in both frames, the flow is dominated by a strong solid-body rotation (along a third axis).
Such large advection speeds are known to cause accuracy issues that can lead to suppression of instabilities \citep[e.g.][]{springel2010}.
The time-step in XSHELLS is adjusted to ensure stability, but we checked accuracy by dividing the time-step by 3 to 7 on several cases.
We found that when instabilities are saturated, stable time steps are small enough to ensure accurate results.
However, we have noticed that for one specific case very close to the onset of instability, the growth rate was biased towards stability.
As not all runs could be checked, it is not impossible that a few such cases are still included in our results, but it would not change the conclusion drawn in this paper.

In the following, we call hydrodynamic cases those with no magnetic field or with magnetic energy lower than $10^{-16}$ to ensure no perturbation of the flow by the magnetic field.

\section{Hydrodynamics}
\subsection{Laminar Base Flow}
\label{base_flow}

The primary flow forced by precession in a sphere is mainly a tilted solid body rotation, a flow of uniform vorticity \citep{poincare1910precession}.
In a spherical container, the direction and amplitude of the fluid rotation vector are governed by a balance between the viscous torque at the core-mantle boundary and the gyroscopic torque resulting from the precession of the liquid core \citep{busse1968steady,Noir2003}.
We briefly recall in section \ref{appendixA1} of appendix \ref{appendixA} the derivation of \citet{busse1968steady}.
A limitation of this general formulation, accounting for the Ekman boundary layer action, arises from the implicit nature of the final equation. Indeed, while approximate expressions can be obtained in certain limits (e.g. \citet{boisson2012earth}), we cannot derive a general analytical explicit solution. In the context of a precessing ellipsoid, \citet{Noir2013} proposed an alternative to the torque balance, using a simpler ad-hoc viscous term.
Using this successful approach in a spherical shell, we obtain an explicit expression of the dimensionless fluid rotation vector $\boldsymbol{\Omega}$ in the frame of precession (see section \ref{appendixA2} of appendix \ref{appendixA} for details):

\begin{eqnarray}
\Omega_x &=& \frac{1}{1+Po} \frac{[\lambda_i+\chi \cos \alpha] \, \chi \sin \alpha }{\chi (\chi +2 \lambda_i \cos \alpha )+|\underline{\lambda}|^2}, \label{eq:sphereXX} \\
\Omega_y &=& -\frac{1}{1+Po} \frac{\chi  \lambda_r \sin (\alpha)}{\chi (\chi +2 \lambda_i \cos \alpha )+|\underline{\lambda}|^2}, \\
\Omega_z &=& \frac{1}{1+Po}  \frac{\chi ( \chi \cos^2 \alpha +2 \lambda_i \cos \alpha)+|\underline{\lambda}|^2}{\chi (\chi +2 \lambda_i \cos \alpha )+|\underline{\lambda}|^2}, \label{eq:sphereZZ}
\end{eqnarray}
where $\chi=Po/\sqrt{E}$. Note that, in the sphere, there is always a single solution for the basic flow $\boldsymbol{\Omega}$, whereas multiple solutions can be obtained in spheroids or ellipsoids \citep{Noir2003,cebron2015bistable,vormann2018numerical}.

In equations (\ref{eq:sphereXX})-(\ref{eq:sphereZZ}), the complex viscous damping coefficient $\underline{\lambda}=\lambda_r+\textrm{i} \lambda_i$ of the spin-over mode is the sum of the (real) viscous damping $\lambda_r$ and the (real) viscous correction to the inviscid frequency $\lambda_i$ of the spin-over mode (\ref{lambdaInv}), such that $|\underline{\lambda}|^2=\lambda_r^2+\lambda_i^2$. For a full sphere, \citet{greenspan1968} derived an expression for $\underline{\lambda}$ at the order $\mathcal{O}(\sqrt[]{E})$. To account for the presence of an inner core, one can use  
\begin{equation}
\underline{\lambda}=\underline{\lambda}_{(\eta=0)}\, \frac{1+\eta^4}{1-\eta^5} \label{eq:dety00}
\end{equation}
for a no-slip inner core \citep{rieutord2001damping}. Moreover, viscous corrections of $\underline{\lambda}$ should be considered for finite values of $E$ (see appendix \ref{appendixA1}). In the following, we account for these various corrections by calculating $\underline{\lambda}$ using equation (\ref{eq:lambda}). 

\begin{figure}[t]
  \begin{center}
	 \includegraphics[width=0.7 \linewidth]{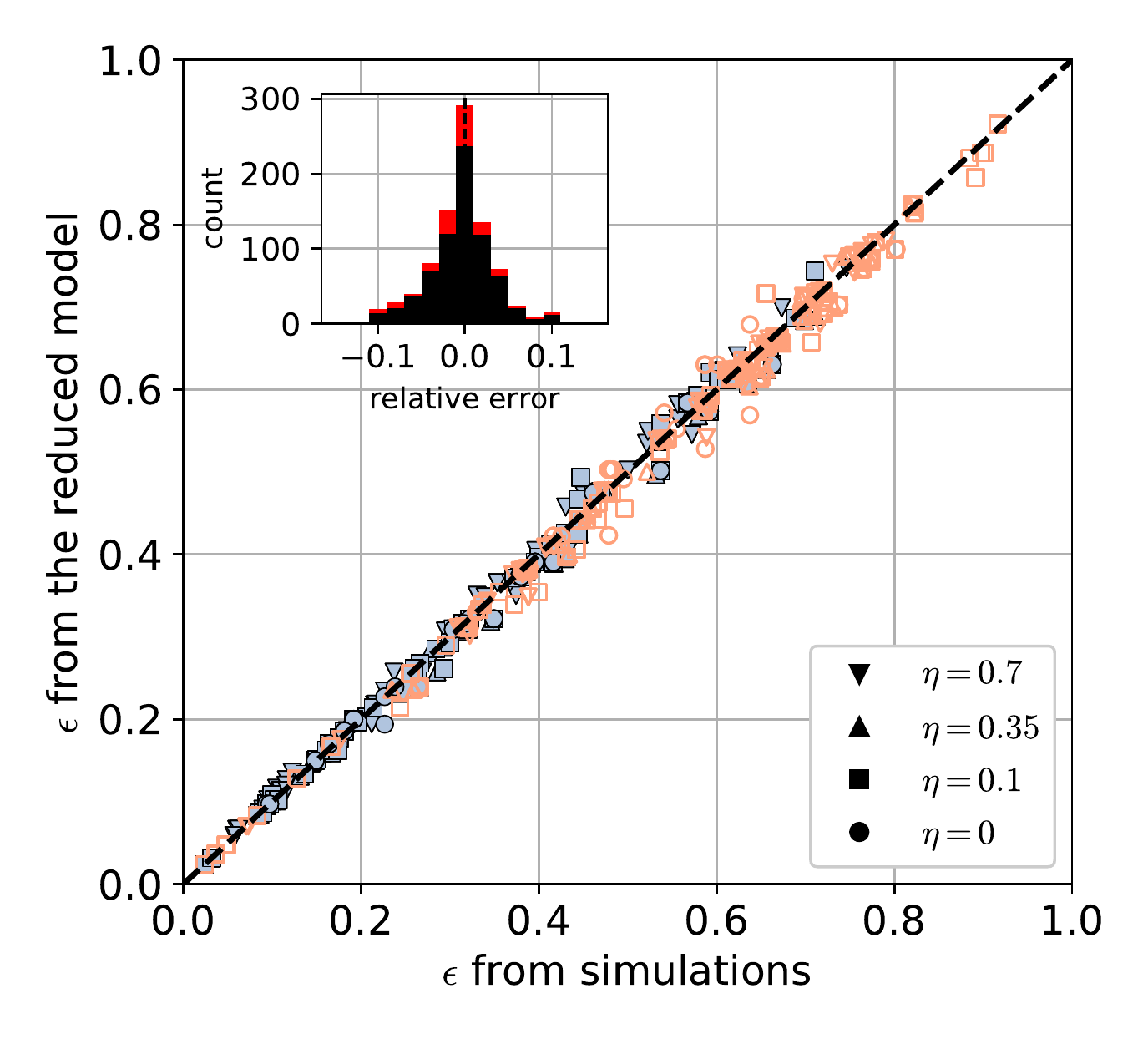}
    \caption{Comparison between the differential rotation $\epsilon$ of the reduced model (eq. \ref{eq:RotDif00} -- dashed line) with all 866 simulations with the required diagnostics (including dynamos).
    Filled and open symbols represent stable and unstable flows, respectively.
    The inset shows the distribution of relative error between reduced model and simulations for all 866 runs (red) and for the 720 ones not affected by magnetic field (black).}
         \label{fig:diffrot}                
  \end{center}
\end{figure}   

Following \cite{lin2015shear}, we note $\epsilon$ the differential rotation $\epsilon=||\boldsymbol{\Omega} - \vect{\tilde{\Omega}_s} ||$ between the fluid and the cavity. From the approximated solution of uniform vorticity (\ref{eq:sphereXX})-(\ref{eq:sphereZZ}) we derive the following analytical expression
\begin{equation}
\epsilon = \left| \frac{ \chi \sin \alpha }{1+Po} \right| \frac{1}{\sqrt{\chi (\chi +2 \lambda_i \cos \alpha )+|\underline{\lambda}|^2}}.\label{eq:RotDif00} 
\end{equation}
The equations and the forcing being centro-symmetric, we calculate the uniform vorticity from the symmetric toroidal energy $E_{s,T}$ of the flow that superimposes on the rotation of the cavity,
\begin{equation}
E_{s,T}= \frac{1}{2} \int \boldsymbol{u}_{s,T}^2 \, \textrm{d}V,
\end{equation}
with $\boldsymbol{u}_{s,T}= [\vect{u}_{m,T}(\vect{r})-\vect{u}_{m,T}(\vect{-r})]/2$ and $\vect{u}_{m,T}=\vect{u}_T- \vect{\Omega_s}  \times \vect{r}$, where $\vect{u}_T$ is the toroidal component of the velocity field such that $\vect{u}_T = \nabla \times (T\vect{r})$.
Thus, the (uniform vorticity) differential rotation is calculated using
\begin{equation}
\epsilon=\sqrt{\frac{2 E_{s,T}}{I_c}}, \label{eq:epsilon}
\end{equation}
where $I_c=8 \pi (1-\eta^5)/15 $ is the moment of inertia (per unit of mass) of the spherical shell enclosing the fluid.
Fig. \ref{fig:diffrot} shows that equation (\ref{eq:RotDif00}) is in quantitative agreement with the uniform vorticity component deduced from the energy in our hydrodynamic simulations, which validates both our reduced model and the estimated differential rotation in the numerics from the toroidal symmetric energy.
However, relative deviations up to 10\% remain between the measured and predicted $\epsilon$ (see error distribution in Fig. \ref{fig:diffrot}).
The distribution of deviations does not change much when keeping only the lowest viscosity or only the stable simulations.
Thus, these deviations are either due to the way we measure the differential rotation, or to the approximation used to obtain equation \ref{eq:RotDif00} (see \S\ref{appendixA2}).
Finer measures of the differential rotation may reduce these deviations, but were unfortunately not available from our runs.
While the origin of these deviations is thus unclear, differential rotation of most cases are accurately predicted by equation \ref{eq:RotDif00} and measured using the toroidal symmetric energy by equation \ref{eq:epsilon}.

\begin{figure}                
	\begin{center}
			 \includegraphics[width=0.7 \linewidth]{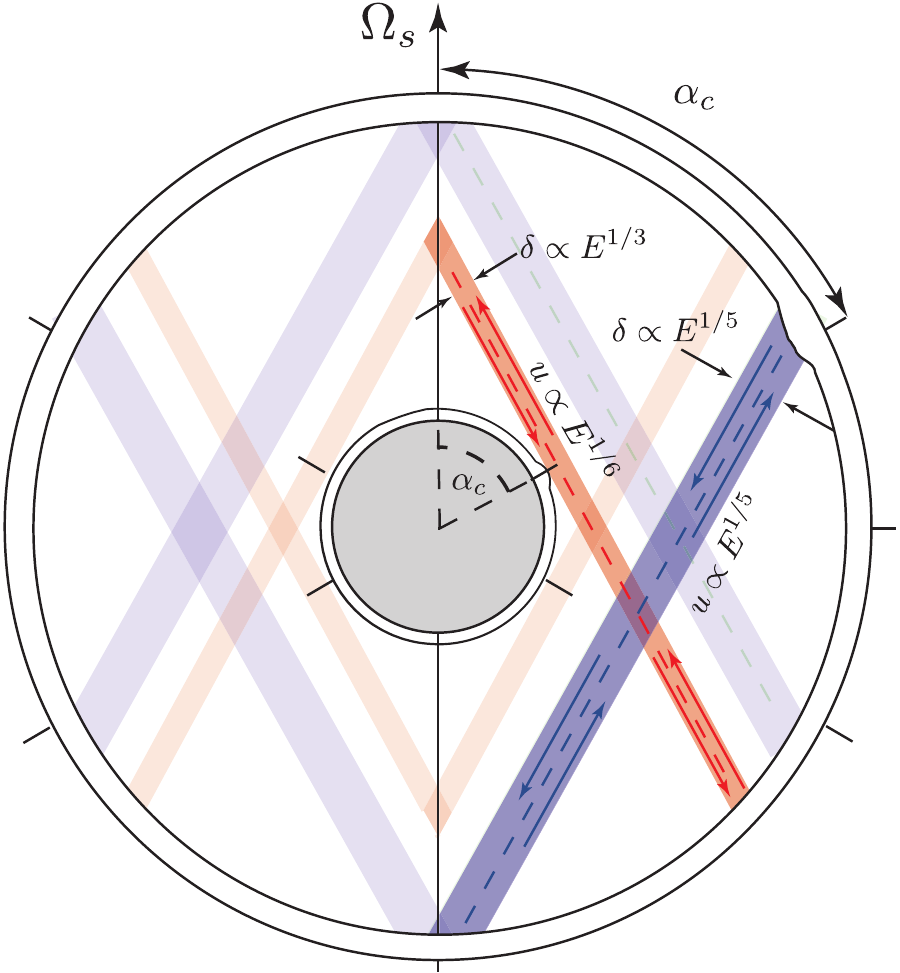} 
		\caption{Schematic of the conical shear layers spawn from the critical colatitude $\alpha_c$ given by $\cos \alpha_c=1/2$.}
		\label{fig:inertialWaves}   
	\end{center}
\end{figure}  

\begin{figure}                
	\begin{center}
		\begin{tabular}{cccc}
			\subfigure[]{\includegraphics[width=0.24 \linewidth]{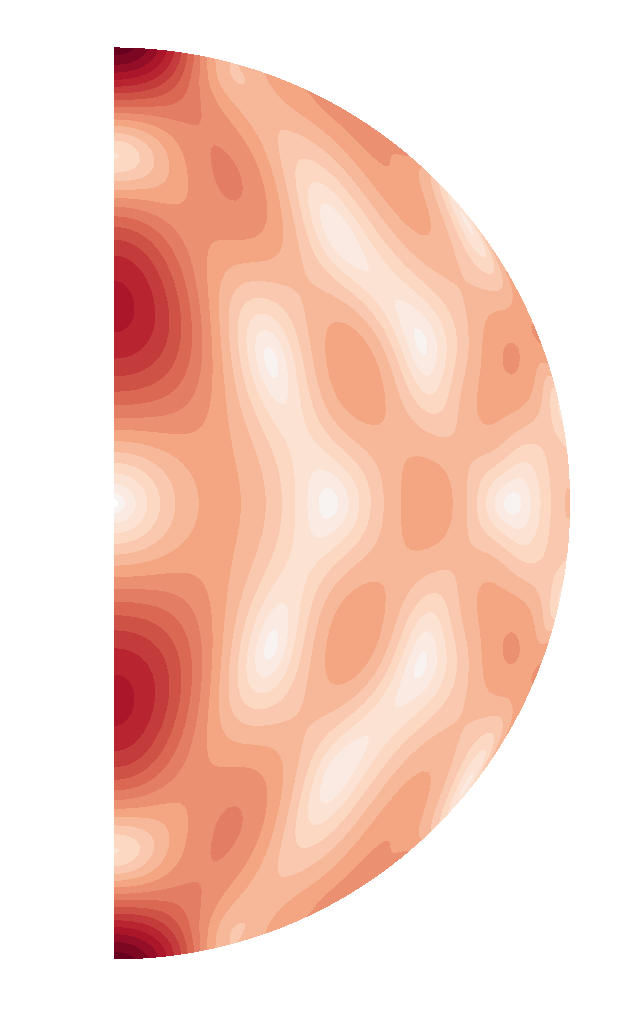}} &
			\subfigure[]{ \includegraphics[width=0.24 \linewidth]{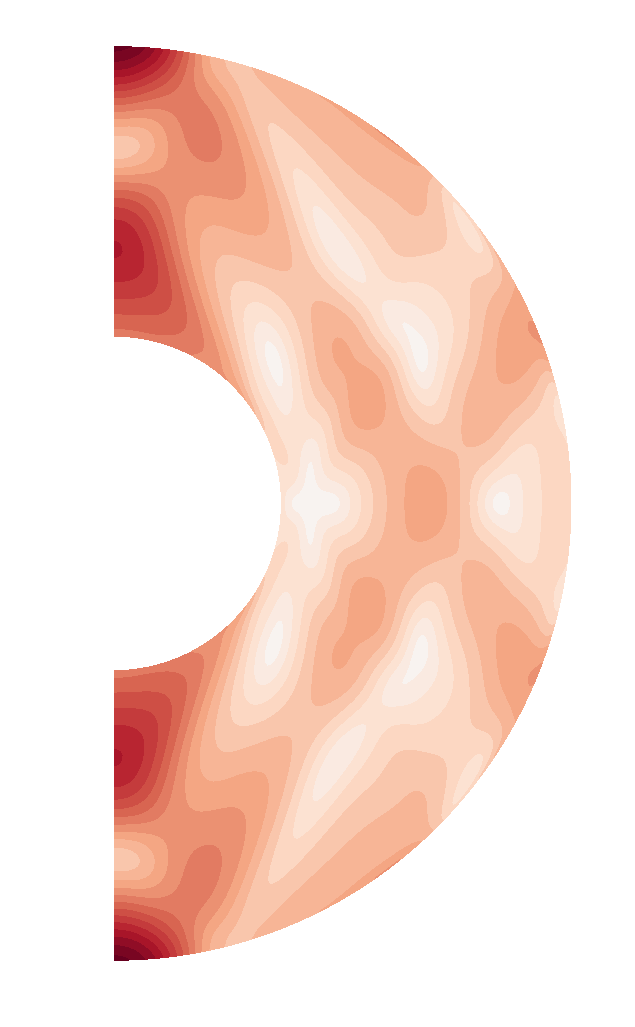}}
			\subfigure[]{\includegraphics[width=0.24 \linewidth]{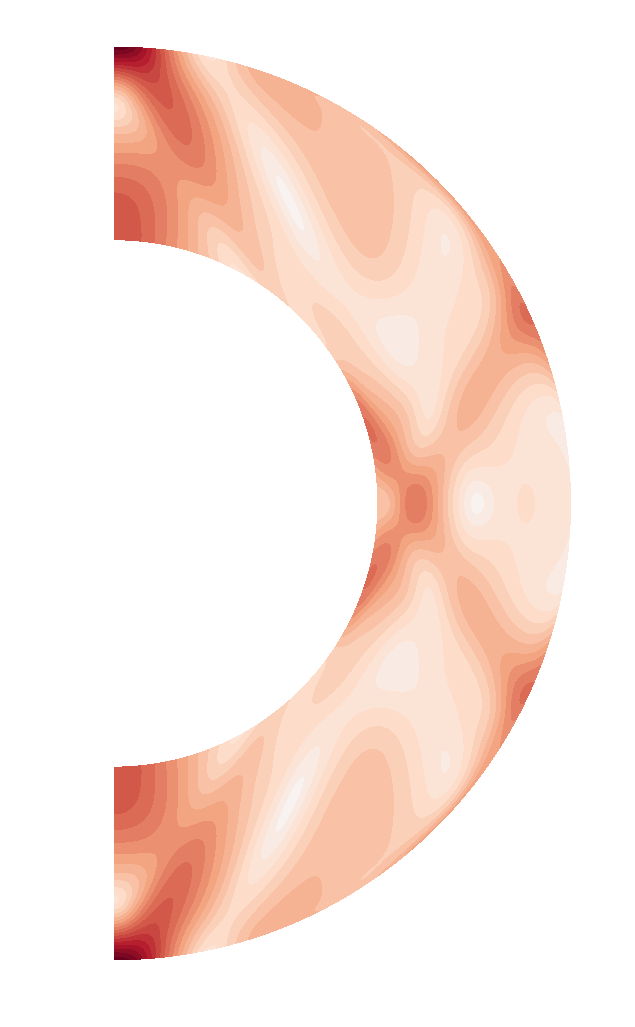}} &
			\subfigure[]{ \includegraphics[width=0.24 \linewidth]{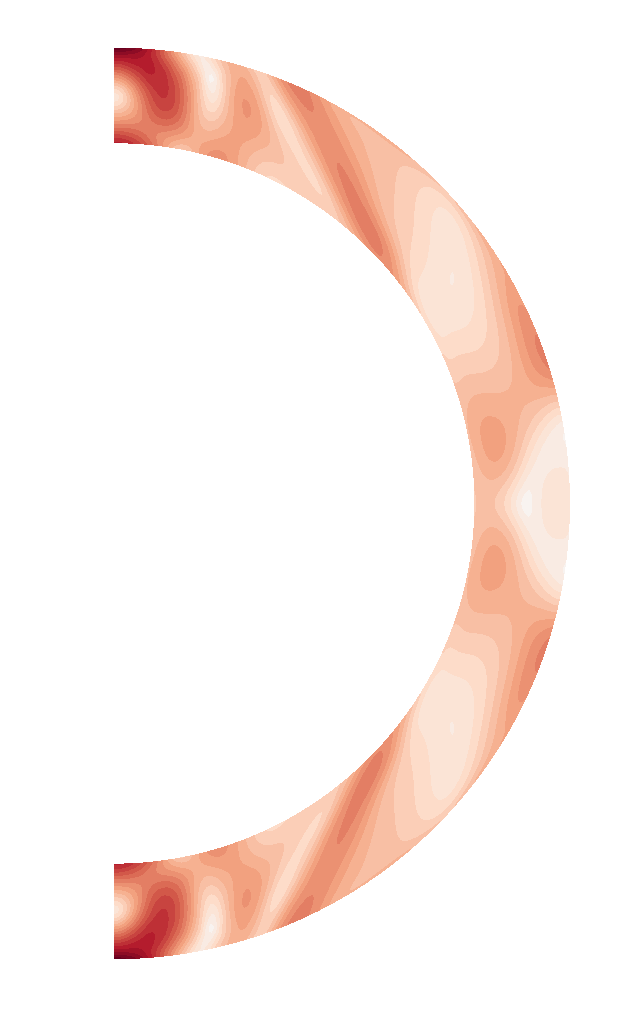}}
		\end{tabular}
		\caption{Contour plot of an instantaneous azimuthal average of the kinetic energy in the fluid frame for (a) $\eta=0.01$, (b) $\eta=0.3$, 
			(c) $\eta=0.5$ and (d) $\eta=0.7$. In each case the flow is stable. $\alpha=120^\circ$, $Po=1\times 10^{-3}$ and $E=3.0\times 10^{-5}$. Contours range from $0$ (white) to $0.028$ (dark red)}.
		\label{fig:mean_E}   
	\end{center}
\end{figure} 

In addition to the uniform vorticity flow, a secondary viscous circulation will develop in the interior due to the Ekman pumping at the ICB and at the CMB. In contrast with the classical uniform thickness of the planar Ekman boundary layer, oscillatory motions in a sphere result in local discontinuities at some critical latitudes, propagating in the interior along cones aligned with the axis of rotation (Fig. \ref{fig:inertialWaves}).
For an oscillatory motion at an angular frequency $\omega$, it corresponds to the colatitude $\alpha_c$ such that $\cos \alpha_c=\omega/2$.
For precession, $\omega=1$, which gives $\alpha_c=60^{\circ}$, i.e. a critical latitude of $30^{\circ}$ \citep{kerswell1995internal,hollerbach1995oscillatory,stewartson1963motion,noir2001numerical}.
Both the ICB and the CMB generate oblique shear layers.
Based on the work of \citet{stewartson1963motion}, \citet{noir2001numerical} corrected the predictions of \citet{kerswell1995internal} and obtained the following scalings for the width and strength of the flow in these oblique shear layers:
\begin{eqnarray}
\delta_{CMB}\propto E^{1/5}, \qquad
U_{CMB}\propto \epsilon E^{1/5} \label{scalingIWvelocity_CMB}.
\end{eqnarray}
and,
\begin{eqnarray}
\delta_{ICB}\propto E^{1/3}, \qquad U_{ICB}\propto  \eta \epsilon E^{1/6}, \label{scalingIWvelocity_ICB} 
\end{eqnarray}
The subscript refers to the region at the origin of the internal structures (Fig. \ref{fig:inertialWaves}).  

To identify these structures in our numerical simulations, we look at the flow from a frame of reference attached to the mean rotation of the fluid. Fig. \ref{fig:mean_E} represents the azimuthal average kinetic energy in a meridian plan for different inner core sizes, clearly exhibiting conical shear layers. The ones spawned from the ICB critical latitude being more intense, they quickly dominate as the inner core radius increases.

\subsection{Hydrodynamic instabilities}\label{hydro_instability}

We track the instability onset by looking for non-zero anti-symmetric energy \citep{lorenzani2001fluid2,lin2015shear}:
\begin{equation}
E_a= \frac{1}{2} \int \boldsymbol{u_a}^2 \, \textrm{d}V, \label{eq:Ea}
\end{equation}
where $\boldsymbol{u_a}=[\boldsymbol{u}(\vect{r})+\boldsymbol{u}(-\vect{r})]/2$. Although centro-symmetric unstable flows exist \citep{hollerbach2013parity}, they are however limited to a narrow range of parameters with moderate to large values of $E$ \citep{lin2015shear}. Disregarding these possible instabilities, we focus on unstable flows which break the centro-symmetry, i.e. with $E_a \neq 0$.

\begin{figure}
\begin{center}
\includegraphics[width=0.8 \linewidth]{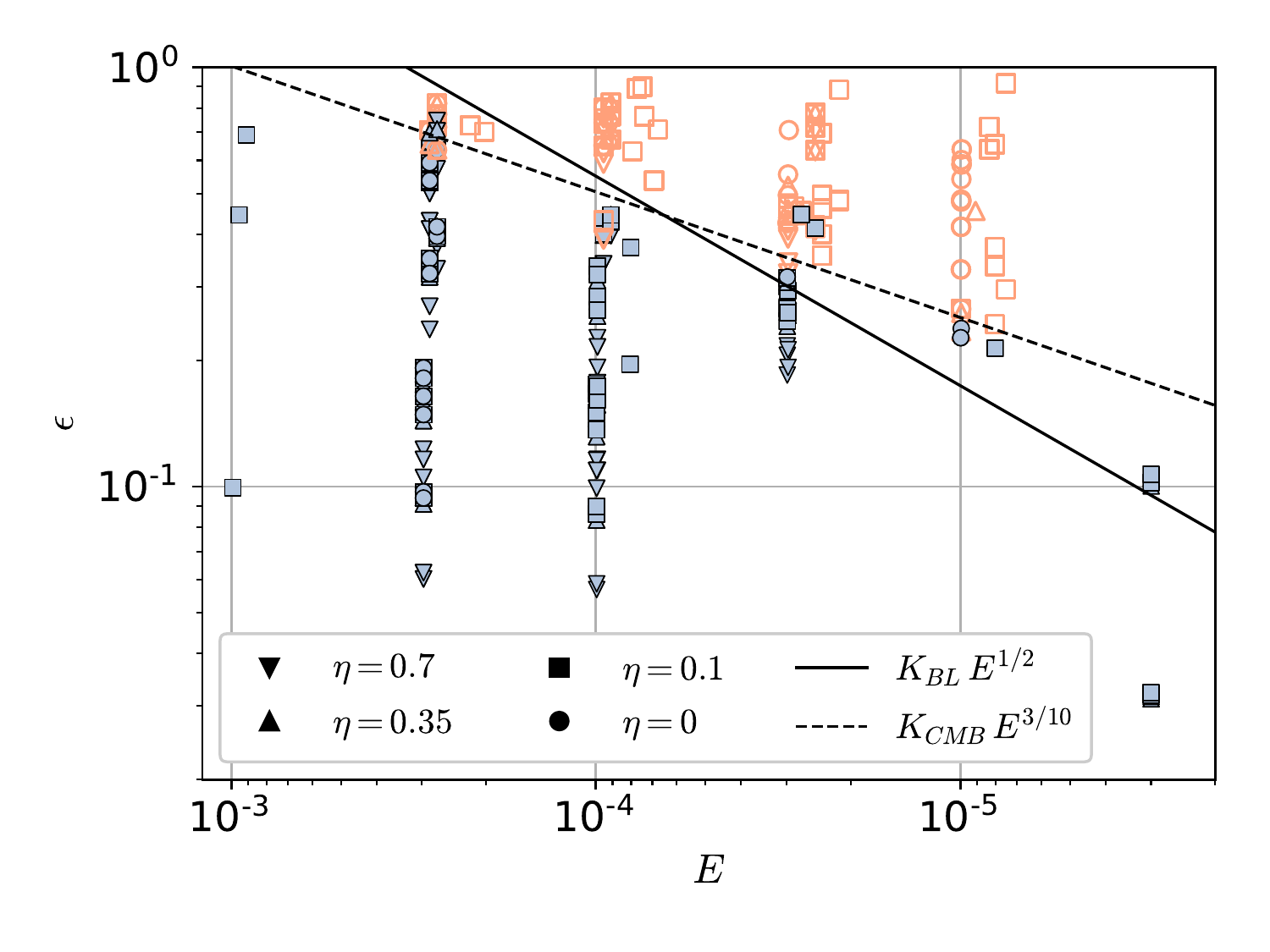}
\includegraphics[width=0.8 \linewidth]{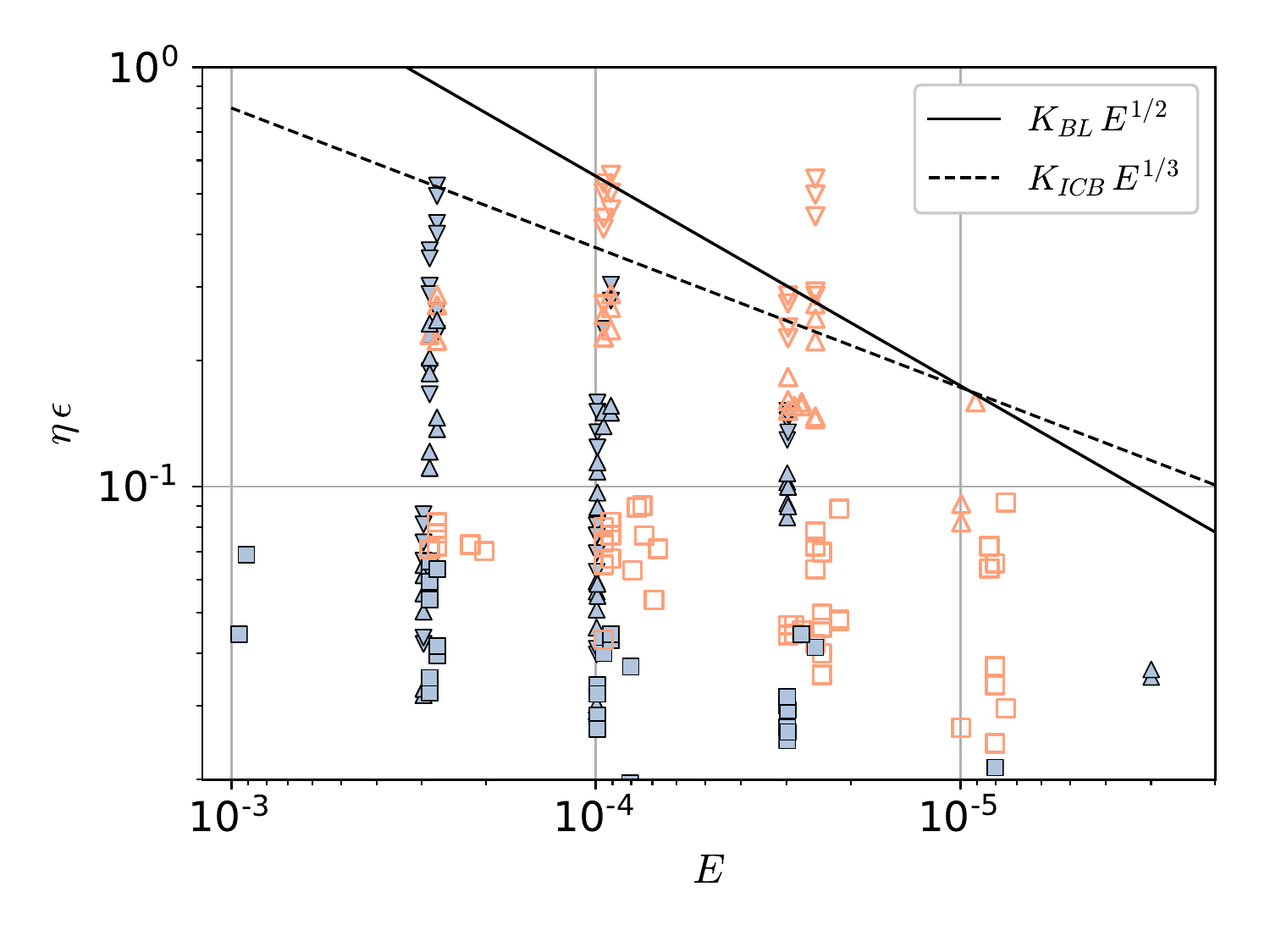}
\caption{Stability criteria, where filled and open symbols represent stable and unstable flows, respectively.
Top: for CMB related instabilities, the dashed and solid lines represent respectively the theoretical CSI (eq. \ref{eq:onsetCMB}) and boundary layer (eq. \ref{eq:BL_CMB}) stability criteria.
Bottom: same criteria for the ICB related instabilities (eq. \ref{eq:onsetICB} and \ref{eq:BL_ICB}).
We have set $K_{CMB}=K_{ICB}=8$ and $K_{BL}=55$.
Anomalous points (ie stable points above the lines and unstable points below) are plotted on top.
Only the 473 points with $Po < 0.3$ and low magnetic energy are shown.
Note that keeping only $Po < 0.1$ removes the two anomalous stable points near $E=3 \times 10^{-5}$, which have been carefully checked to be stable even when the time-step is reduced (see \S\ref{sec:num}).}
\label{fig:stabdiagCSI}
\end{center}
\end{figure}

\begin{figure}
\begin{center}
\includegraphics[width=0.7 \linewidth]{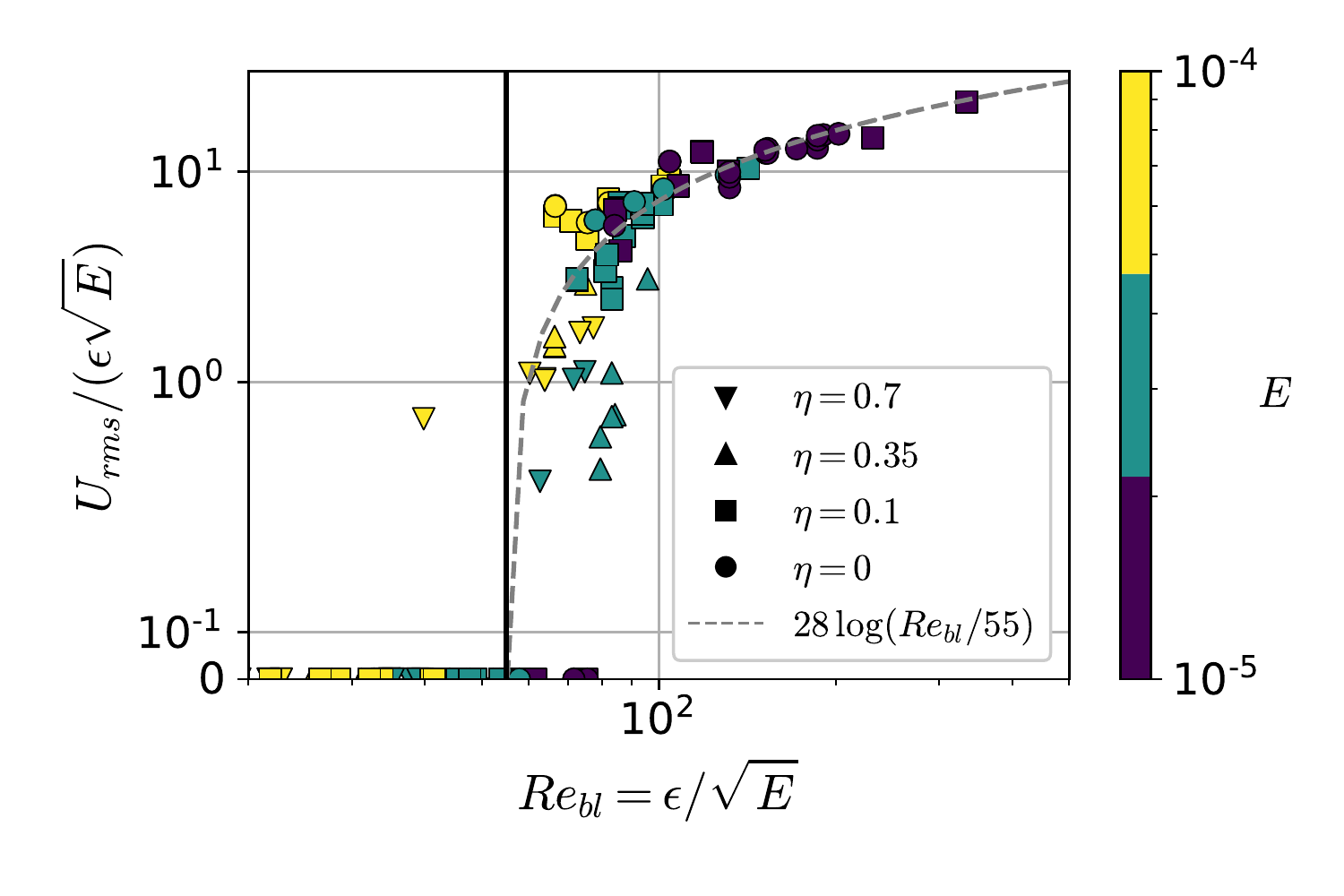}
\includegraphics[width=0.7 \linewidth]{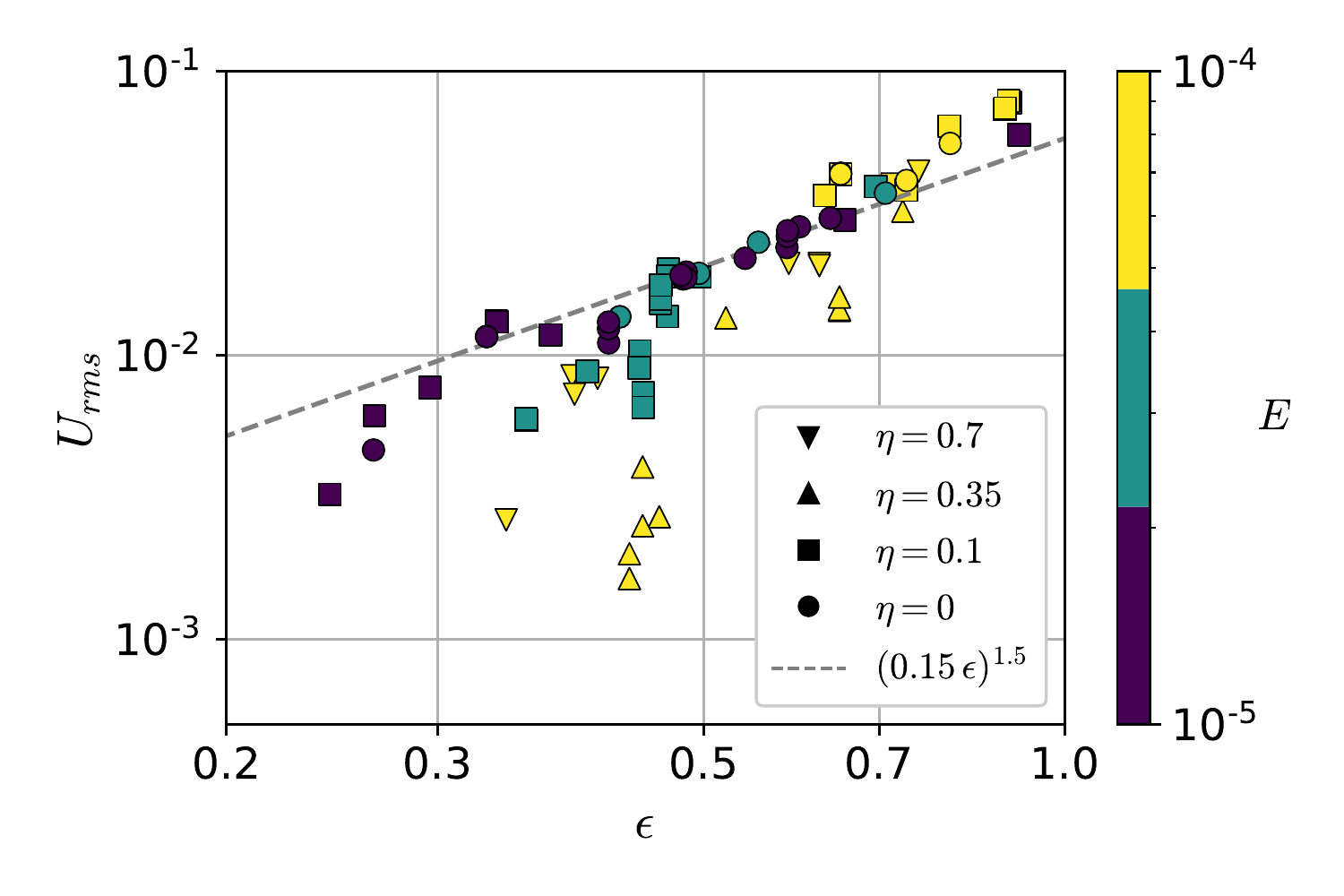}

\caption{Two ways to collapse the turbulent fluctuation velocity $U_{rms} = \sqrt{4 E_a / V}$ of the 84 simulations with $Po < 0.1$, $E \leq 10^{-4}$ and low magnetic energy.
The dashed lines are tentative and speculative laws that explain some of our data; they are not supported by theory and should not be used for extrapolation.
}
\label{fig:EnergyAntiSym}
\end{center}
\end{figure}

It has been recently argued by \citet{lin2015shear} and \citet{lorenzani2001fluid2} that the oscillating conical shear layers, originating from the CMB Ekman boundary layers, can couple non-linearly with two inertial modes,  $\vect{u}_1$ and $\vect{u}_2$, leading to a parametric resonance, the so-called Conical Shear Instability \citep[CSI,][]{lin2015shear}.
For precession, the two free inertial modes are subject to the following selective rules \citep{kerswell2002elliptical}:
\begin{eqnarray}
\omega_1\pm\omega_2&=&1,\cr 
m_1\pm m_2&=&1,\cr 
l_1\pm l_2&=&1.
\end{eqnarray}
where $\omega_{1,2}$ and $m_{1,2}$ are the frequencies and azimuthal wave numbers of the two free inertial modes, respectively. $l_{1,2}$ is the degree of the Legendre polynomial characterizing the latitudinal complexity. 

Based on the scaling of the oblique shear layers emanating from the CMB (Fig. \ref{fig:inertialWaves}), \citet{lin2015shear} proposed that the onset of the CSI is governed by a critical value of the differential rotation $\epsilon$, scaling as
\begin{equation}
\epsilon_c \propto E^{3/10} ,
\end{equation}
in agreement with their numerical simulations in a full sphere as well as the experimental results from \citet{Goto2014}. The same argument can be used in the spherical shell to derive a criteria for the onset of a CSI driven by the oblique shear layers emanating from the CMB as well as from the ICB.

For clarity, we name CSI-ICB and CSI-CMB, the parametric instabilities of the conical shear layers spawned from the Ekman boundary layer of the ICB and CMB, respectively. Adopting the same approach as \citet{lin2015shear}, with the scaling of shear layers shown in Fig. \ref{fig:inertialWaves}, we obtain
\begin{equation}
\epsilon_c=K_{CMB} E^{3/10},
\label{eq:onsetCMB}
\end{equation}
for the CSI-CMB and 
\begin{equation}
\epsilon_c=K_{ICB} \frac{E^{1/3}}{\eta}
\label{eq:onsetICB}
\end{equation}
for the CSI-ICB, where $K_{CMB}$, $K_{ICB}$ are two constants.

In addition, the CMB and ICB Ekman boundary layers may be unstable to a local shear instability.
The onset of this boundary-layer instability is characterized by the local Reynolds number $Re_{bl} = v \delta / \nu \approx 55$ \citep[e.g.][]{lorenzani2001fluid2,sous2013friction} based on the Ekman layer thickness $\delta=\sqrt{\nu/\Omega} = \sqrt{E}R$ and the maximum (differential) tangential velocity $v$ at the edge of the boundary layers, with $v=\epsilon$ at the CMB and $v = \eta \epsilon$ at the ICB.
The stability criteria for the ICB and CMB read
\begin{eqnarray}
Re_{CMB}&=&\frac{\epsilon}{\sqrt{E}} >K_{BL}, \label{eq:BL_CMB}\\
Re_{ICB}&=&\frac{ \eta \epsilon}{\sqrt{E}} >K_{BL},
\label{eq:BL_ICB}
\end{eqnarray}
with $K_{BL} \approx 55$ \citep{lorenzani2001fluid2,sous2013friction}, and the associated instabilities will be noted respectively BL-CMB and BL-ICB. In all cases, the outer boundary will become unstable first.

We arbitrarily distinguish the stable and unstable cases by $E_a<10^{-10}$ and $E_a>10^{-10}$, respectively. To unravel the underlying destabilizing mechanism, we represent our results in an ($\epsilon,E$) parameter space against the above mentioned onset criterion for the CSI-CMB and CSI-ICB (Fig. \ref{fig:stabdiagCSI}).
We cover a wide range of parameters, $\eta=0, 0.1, 0.35, 0.7$, $Po<0.3$ and $E \leq 10^{-3}$.
Contrarily to the CMB-related criteria, the ICB-related ones do not separate stable from unstable points, and we thus conclude that the first instability is due to the CMB.
Fig. \ref{fig:stabdiagCSI} further suggests that in our numerical simulations the first instability is a CSI-CMB at moderate to large Ekman numbers.
Meanwhile, below $E=3\times10^{-5}$, our theory predicts that the boundary layer is unstable before the CSI-CMB.
One can notice the robustness of the CMB-related instability criteria for all inner core radii investigated up to $\eta=0.7$.

In order to distinguish clearly between the two mechanisms one should carry out numerical simulations in the range $10^{-8}<E<10^{-6}$, a range of values still hardly accessible. Further considerations regarding the prevalence of a possible viscous boundary layer at the ICB will be discussed in the last section in a geophysical context at very low Ekman numbers.

As a proxy for the turbulent fluctuation level around the mean flow, let us define the root-mean-square fluctuation velocity as $U_{rms} = \sqrt{4 E_a/V}$, where $V=4/3\pi (1-\eta^3) $ is the fluid volume and the factor $4$ assumes equipartition between symmetric and anti-symmetric turbulent fluctuations.
While the onset of instability is reasonably well captured by the BL-CMB at low viscosity ($E \lesssim 3 \times 10^{-5}$), it is more difficult to understand the amplitude of the saturated turbulent velocity $U_{rms}$.
Figure \ref{fig:EnergyAntiSym} shows two ways we found to collapse our data, including a viscosity-free law (fig. \ref{fig:EnergyAntiSym}b).
None of them is fully convincing and further calculations at lower Ekman numbers are clearly necessary to uncover a saturation scaling law.
Nevertheless, some systematic behavior is captured in this figure with rather low amplitude fluctuations $\epsilon \sqrt{E} \lesssim U_{rms} < \epsilon/10$ in the planetary parameter range $\epsilon \ll 1$, $E \lll 1$.

\subsection{Instability flow structure} 

\begin{figure}[t]
  \begin{center}
  	\begin{tabular}{ccc}
  \subfigure[]{\includegraphics[width=0.3 \textwidth]{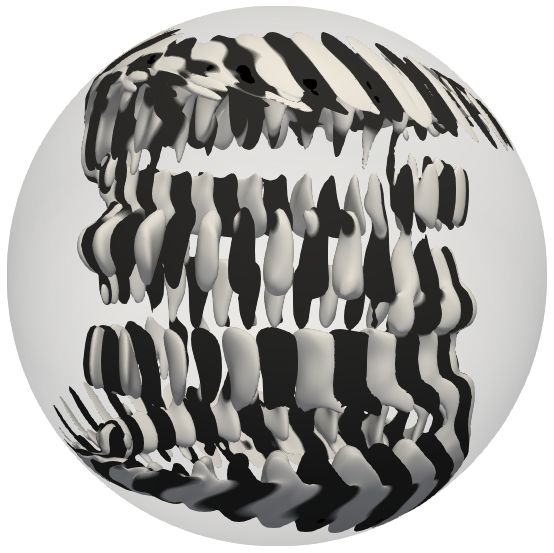}} &
   \subfigure[]{ \includegraphics[width=0.3 \textwidth]{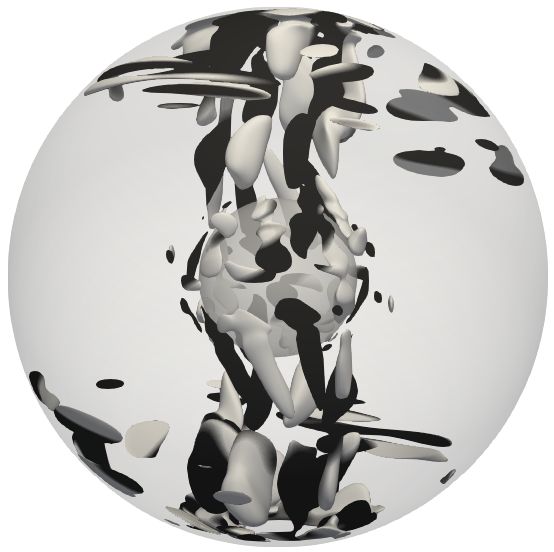}}
   &
   \subfigure[]{ \includegraphics[width=0.3 \textwidth]{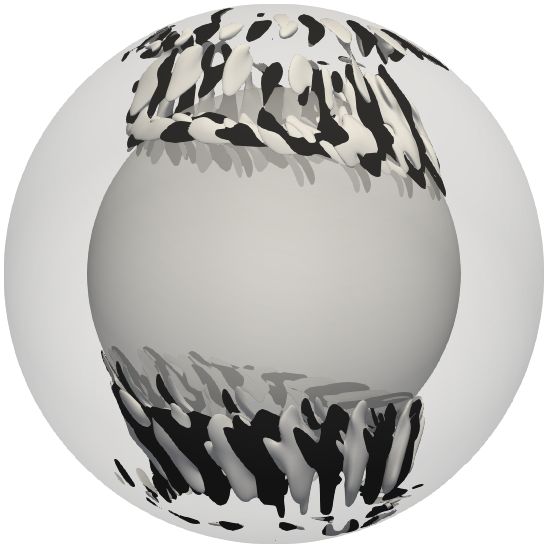}}\\
\end{tabular}
 \caption{Isosurfaces of the  anti-symmetric energy in the mean fluid rotating frame frame for (a) $\eta=0.01$, $Po=8\times 10^{-3}$ (b) $\eta=0.3$, $Po=8\times 10^{-3}$ and (c) $\eta=0.7$, $Po=8.5\times 10^{-3}$. In each case $\alpha=120^\circ$ and $E=3.0\times 10^{-5}$. Color correspond to positive (black) and negative (white) axial velocity. The snapshots are taken during the initial growth phase of the instability.}
  \label{fig:Easym_3D}         
   \end{center}
\end{figure}

Fig. \ref{fig:Easym_3D} represents the axial velocity during the growth phase of the instability in the system for three different inner core sizes ($\eta=0.01,0.3,0.7$) and for the following control parameters, $\alpha=120^\circ$, $E=3.0\times 10^{-5}$, $Po=8\times 10^{-3}$ except for the largest inner core for which the instability is detected only for $Po=8.5 \times 10^{-3}$. Not surprisingly, the smallest inner core (Fig. \ref{fig:Easym_3D}a) is comparable to the full sphere case of \citet{lin2015shear} with two inertial modes of wave numbers $m=17$ and $m=18$ developing in the outer part of the fluid domain. As we increase the volume of the inner core, the modes involved in the parametric resonance remains high order near onset (Fig. \ref{fig:Easym_3D}b and c) and tend to develop in regions above and below the inner core, again exhibiting pairs of inertial modes satisfying the parametric resonant conditions. These results near onset suggest that a CSI-CMB mechanism is operating at this low value of the Ekman number.

\begin{figure*}                
  \begin{center}
  	\begin{tabular}{c}
           \subfigure[]{\includegraphics[width=  0.85\textwidth]{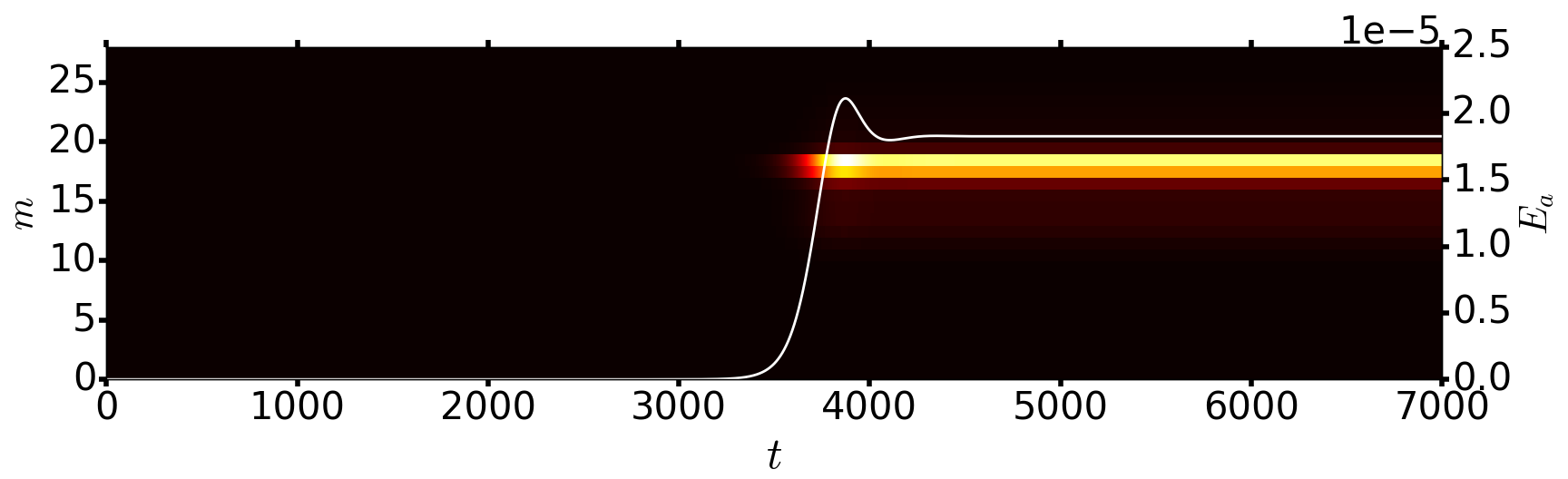}}\\   
           \subfigure[]{\includegraphics[width=  0.85\textwidth]{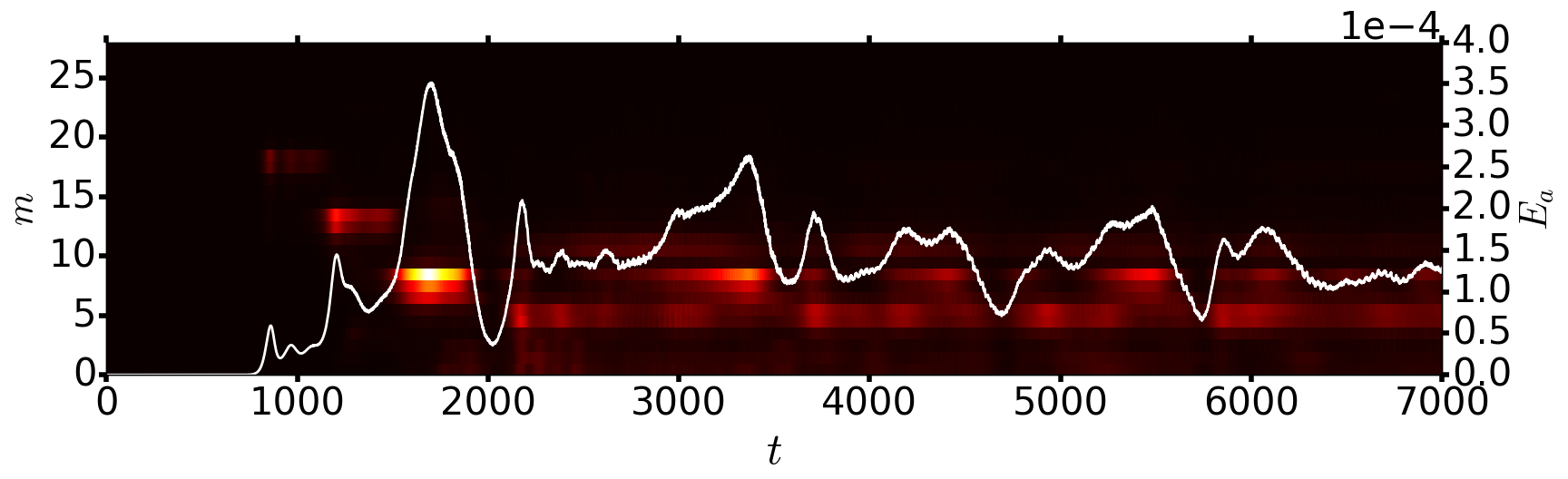}}\\
           \subfigure[]{\includegraphics[width=  0.85\textwidth]{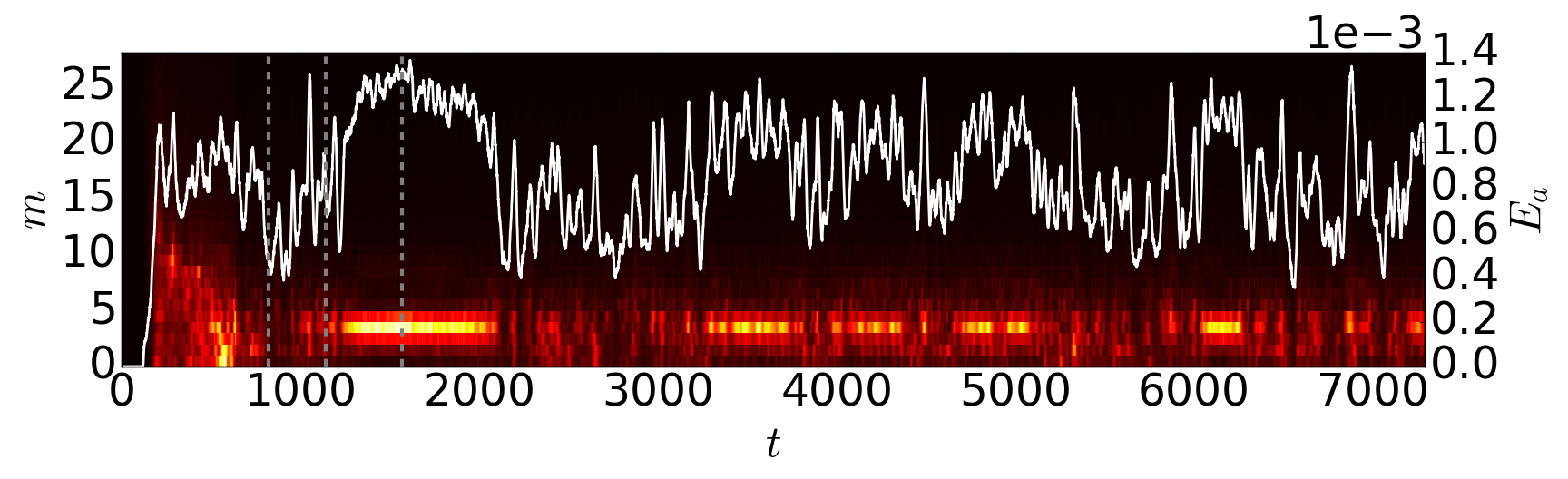}}\\
  \end{tabular}
 \caption{Time evolution of the anti-symmetric kinetic energy of each azimuthal mode $m$ in the fluid frame (color map) and of the total anti-symmetric kinetic energy (White solid line). Fixed $\eta=0.01$, $\alpha=120^\circ$ and $E=3.0\times 10^{-5}$. The Poincar\'e number increases from (a) $Po=7\times 10^{-3}$, (b)  $Po=8.5\times 10^{-3}$ and (c) $Po=1.3\times 10^{-2}$, with dashed gray lines corresponding to the times of figure \ref{fig:LSV001}.}
 \label{fig:EvsTime} 
   \end{center}
\end{figure*}  

Fig. \ref{fig:EvsTime} shows the total anti-symmetric kinetic energy and the anti-symmetric kinetic energy for each azimuthal wave numbers $m$ for increasing Poincar\'e numbers, from $Po=7\times10^{-3}$ just above onset to about 2 times critical at $Po=1.3\times10^{-2}$ for $\eta=0.01$. Near onset, the parametric instability remains at a saturated state (Fig. \ref{fig:EvsTime}(a)).
As shown by Fig. \ref{fig:EvsTime}(b), a very small increase of the forcing at $Po=8.5\times10^{-3}$ leads to a quasi periodic behaviour of the system with a typical period of order $T=500$ \cite[this behaviour is reminiscent of the resonant collapses observed e.g. in][]{lin2014experimental}. Despite the modest increase in $Po$, we observe an anti-symmetric energy at saturation that is an order of magnitude larger, yet pairs of modes in parametric resonance can be clearly identified supporting a CSI-CMB underlying mechanism.

\begin{figure*}
  \begin{center}
  	\begin{tabular}{ccc}
   \subfigure[]{\includegraphics[width=0.3\textwidth]{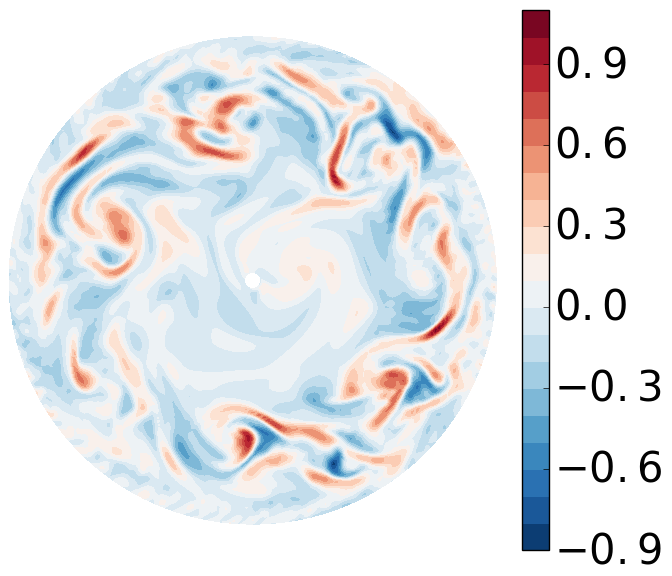}} &
   \subfigure[]{\includegraphics[width=0.3\textwidth]{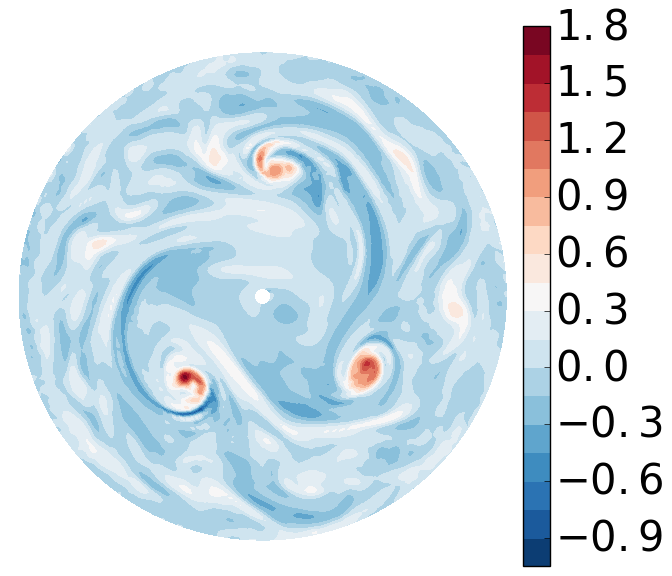}} &
   \subfigure[]{\includegraphics[width=0.325\textwidth]{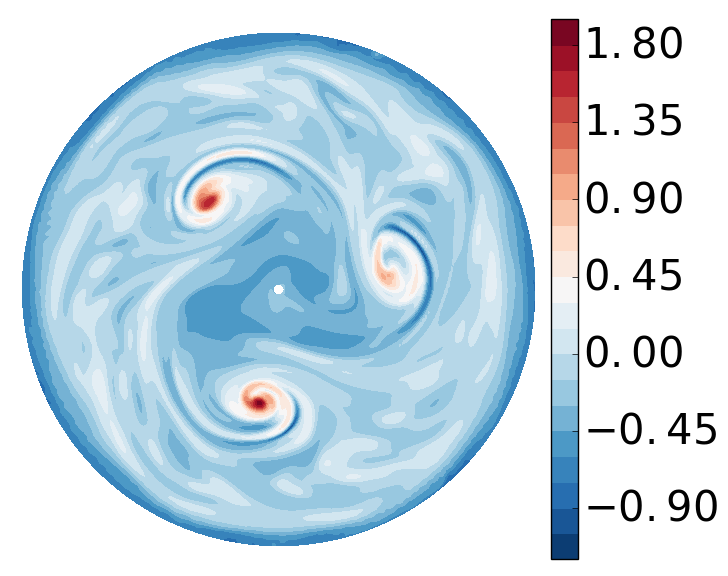}}
 \end{tabular}
 \caption{Snapshot of an equatorial cross-section of the axial vorticity $\omega_z$ in the fluid frame for $\eta=0.01$, $Po=1.3\times 10^{-2}$, $\alpha=120^\circ$ and $E=3.0\times 10^{-5}$, same as Fig. \ref{fig:EvsTime}. (a) $t=822$, (b) $t=1142$ and (c) $t=1567$}
 \label{fig:LSV001} 
   \end{center}
\end{figure*}

In Fig. \ref{fig:EvsTime}(c), we further increase the precession rate to $Po=1.3\times 10^{-2}$. We do not observe a clear initial growth of any particular modes but intermittent states with chaotic and quasi steady phases appear.
At $Po=1.3\times 10^{-2}$ the dynamic of system is no longer quasi periodic in time, rapid fluctuations are observed together with periods of stable energy of modes with $m=3$ as for instance between $t\sim1250$ and $t\sim 2100$.
The system alternates between phases in which $E_a$ is concentrated in the $m=3$ azimuthal mode, contrasting with phases during which $E_a$ is distributed over a wider range of $m$.
In Fig. \ref{fig:LSV001}, we show that three cyclonic large-scale vortices (LSV) are seen in the phases where $E_a$ is concentrated in $m=3$, while they are absent in the other phases.

Finally, we increase $\eta$ from 0.3 to 0.7 at $E=3.0\times 10^{-5}$, $Po=8.5\times 10^{-3}$, $Po=1.3\times10^{-2}$ and $Po=2\times10^{-2}$, to investigate the influence of the inner core on the dynamics above the onset. At $Po=8.5 \times 10^{-3}$ we observe a similar dynamics as for $\eta=0.01$ with quasi-periods increasing with $\eta$, the mode structures are qualitatively similar to the full sphere as seen on Fig. \ref{fig:Easym_3D}.  While at $\eta=0.1,0.3$ we could still observe LSV, although with a shorter life time, they completely disappear for $\eta>0.3$ for $Po=1.3\times10^{-2}$ and $Po=2\times10^{-2}$.
At this Poincar\'e numbers with moderate to large inner cores, the flow exhibits small scale structures with rapid temporal variations.
In contrast with the mode-coupling regime, the typical time scale of the energy fluctuations decreases with increasing inner core size.
These calculations quickly become computationally challenging as the Poincar\'e number is increased.
Since an extensive survey of the parameter space to characterise the onset of the LSV is beyond the scope of this paper, we did not explore the higher $Po$ range where the LSV may be driven even in the presence of a large inner core.

\subsection{Energy Dissipation} \label{sec:dissipa}

\begin{figure}[t]
\begin{center}
\includegraphics[width=0.9 \linewidth]{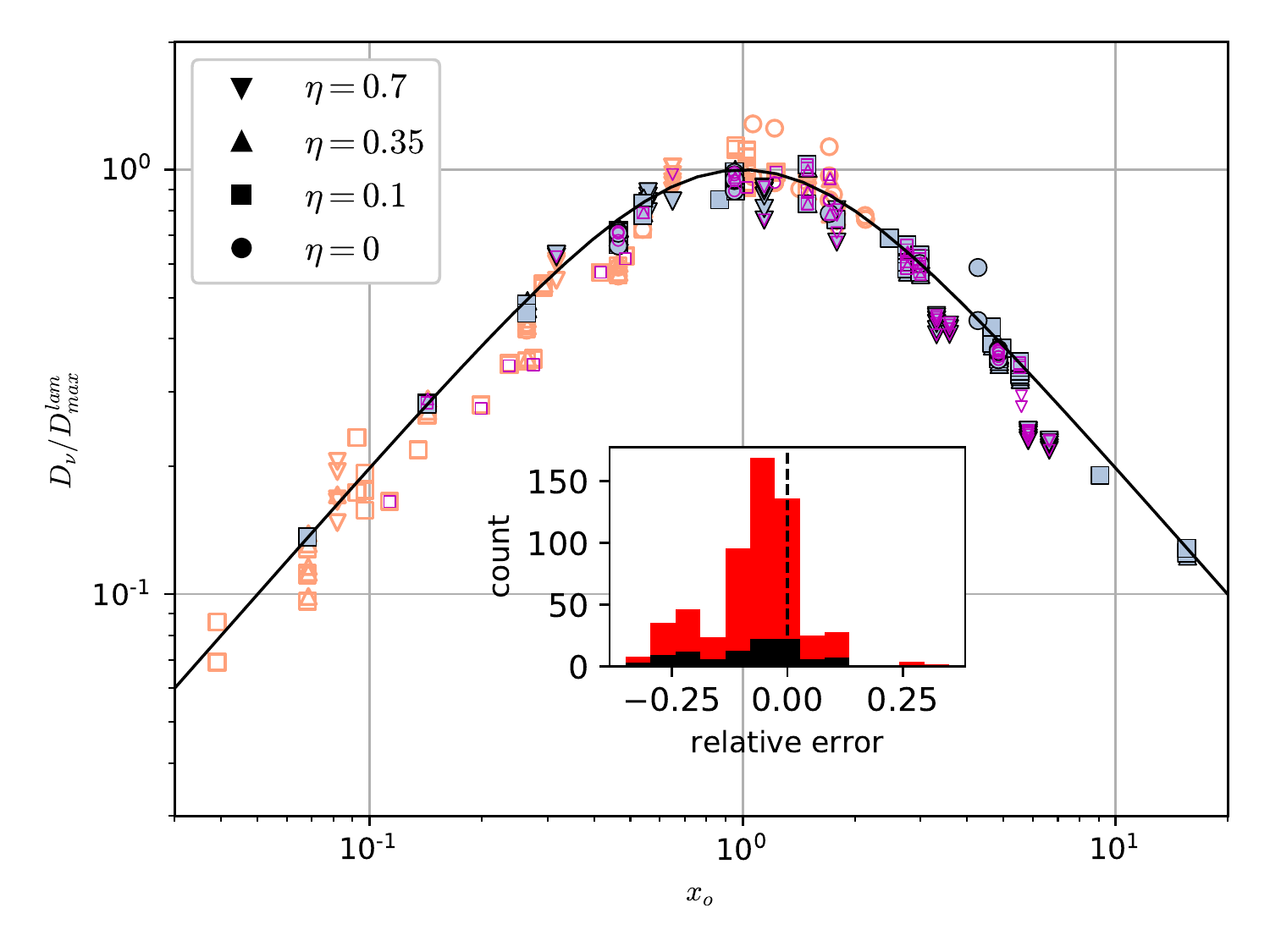}
	\caption{Total viscous dissipation $D_\nu$ obtained in simulations (symbols) and laminar dissipation $D_\nu^{lam}$ (solid line) predicted by equation (\ref{eq:Dnutheo}), as a function of parameter $x_o \simeq 2.62 \, E^{1/2} /Po$.
	Filled and open symbols represent respectively stable and unstable flows for the 573 cases with $Po<0.3$ and $E \leq 10^{-3}$, including MHD runs (magenta symbols, magnetic energy $Em > 10^{-16}$).
	The inset shows the relative difference between $D_\nu$ and $D_\nu^{lam}$ (with the MHD runs in black).}
	\label{fig:energyMax}
\end{center}
\end{figure}

The total viscous dissipation is given by
\begin{equation}\label{eq:numDissip}
D_{\nu}=E \int (\nabla \times \vect{u_m})^2\mathrm{d}V
\end{equation}
with $\vect{u_m}=\vect{u} - \vect{\Omega_s}  \times \vect{r}$ the velocity field in the mantle frame such that $\vect{u_m}=0$ at the boundaries.
This dissipation arises, in absence of instability, purely through viscous friction in the boundary layers at the inner and outer walls and in the oblique shear layers in the bulk.

First, we consider the oblique shear layers in the bulk.
To calculate the associated dissipation, one can restrict the volume integral to the shear layers in equation (\ref{eq:numDissip}).
For the conical shear layer originating from the CMB we obtain $D_{\nu} \sim \epsilon^2 E^{6/5}$ using the scalings (\ref{scalingIWvelocity_CMB}), whereas for the one originating from the ICB we find $D_{\nu} \sim \epsilon^2 \eta^2 E$ using the scalings (\ref{scalingIWvelocity_ICB}).
The viscous dissipation due to boundary friction is given by $D_{\nu}^{BL}=\boldsymbol{\Gamma}_{\nu} \cdot (\vect{\Omega}-\vect{\tilde{\Omega}_s})$ where $\boldsymbol{\Gamma}_{\nu}$ is the associated viscous torque.
Having shown in section \ref{base_flow} that the reduced model performs equally well than the model of \cite{busse1968steady} but provides explicit expression for $\epsilon$, we use equation (\ref{addhocvisc}) for $\boldsymbol{\Gamma}_{\nu}$ to obtain the laminar dissipation
\begin{eqnarray}
D_{\nu}^{BL,lam} = -I_c\, \epsilon^2\,  \lambda_r\, \sqrt{E}. \label{eq:Dnusimple}
\end{eqnarray}
Note that $D_{\nu}^{BL,lam}$ only differs by a factor $\Omega^{1/2}$ (with $\Omega \lesssim 1$, see appendix \ref{appendixA}) from the one obtained with the viscous torque (\ref{eqNoSUP}) of the \cite{busse1968steady} model.
For small Ekman numbers $E \ll 1$, the dissipation in the oblique shear layers is thus negligible, and the total laminar dissipation $D_{\nu}^{lam}$ reduces to $D_{\nu}^{BL,lam}$.
Substituting $\epsilon$ with its expression (\ref{eq:RotDif00}) into equation (\ref{eq:Dnusimple}) leads to
\begin{eqnarray}
D_{\nu}^{lam} = \frac{8 \pi(1-\eta^5) |\lambda_r|\, \sqrt{E}}{15 |1+Po|^2}\,  \frac{|\chi \sin \alpha |^2}{|\chi (\chi +2 \lambda_i \cos \alpha )+\lambda_r^2+\lambda_i^2|}. \label{eq:dnu_red0}
\end{eqnarray}
Defining $x_o=|\underline{\lambda}| E^{1/2}/Po = |\underline{\lambda}|/\chi$ and noticing that the contribution of $\lambda_i$ is always negligible, expression (\ref{eq:dnu_red0}) can be rewritten as
\begin{eqnarray}
D_{\nu}^{lam}=\frac{2 x_o}{1+x_o^2}\, D_{max}^{lam},  \label{eq:Dnutheo}
\end{eqnarray}
with
\begin{eqnarray}
D_{max}^{lam}=\frac{4 \pi}{15}\, \frac{(1-\eta^5) |\lambda_r| Po\,  \sin^2 \alpha}{(1+Po)^2 |\underline{\lambda}|}. 
\end{eqnarray}
$D_{max}^{lam}$ is the maximum laminar non-dimensional viscous dissipation obtained for $x_o \simeq 1$, which is independent of $E$.

Fig. \ref{fig:energyMax} presents $D_{\nu}/D_{max}^{lam}$ from our numerical simulations together with the laminar estimate (\ref{eq:Dnutheo}).
It shows that the laminar boundary-layer dissipation given by the explicit expression (\ref{eq:Dnutheo}) captures the main contribution to viscous dissipation in our simulations, with less than $30\%$ error.
In particular, the maximum viscous dissipation is reached in our low Ekman and low $Po$ simulations for $x_o \simeq 1$ that is $Po \simeq 2.6 \, E^{1/2}$.
Note that the current Earth would be at $x_o \sim 1$, i.e. near the maximum of dissipation, whereas the current Moon would be far from this maximum, at $x_o \sim 10^{-3}$.

Beyond the laminar boundary dissipation, we can compare our results with estimates of turbulent dissipation.
\citet{kerswell1996upper} derived an upper bound for viscous dissipation which reads in our notations:
\begin{eqnarray}
D_{\nu} (1+Po)^3 \leq \left\{
 \begin{array}{ccc}
      8 \pi \chi^2/525 \quad \textrm{for} \, \chi = Po/\sqrt{E} \lesssim 3 \\
      0.43 \quad \textrm{for} \, \chi = Po/\sqrt{E} \gtrsim 3 \\
   \end{array}
   \right. \label{eq:k96}
\end{eqnarray}
All our simulations have a dissipation lower than this upper bound, sometimes several order of magnitude lower, even for unstable flows.

\begin{figure}
\begin{center}
\includegraphics[width=0.8 \linewidth]{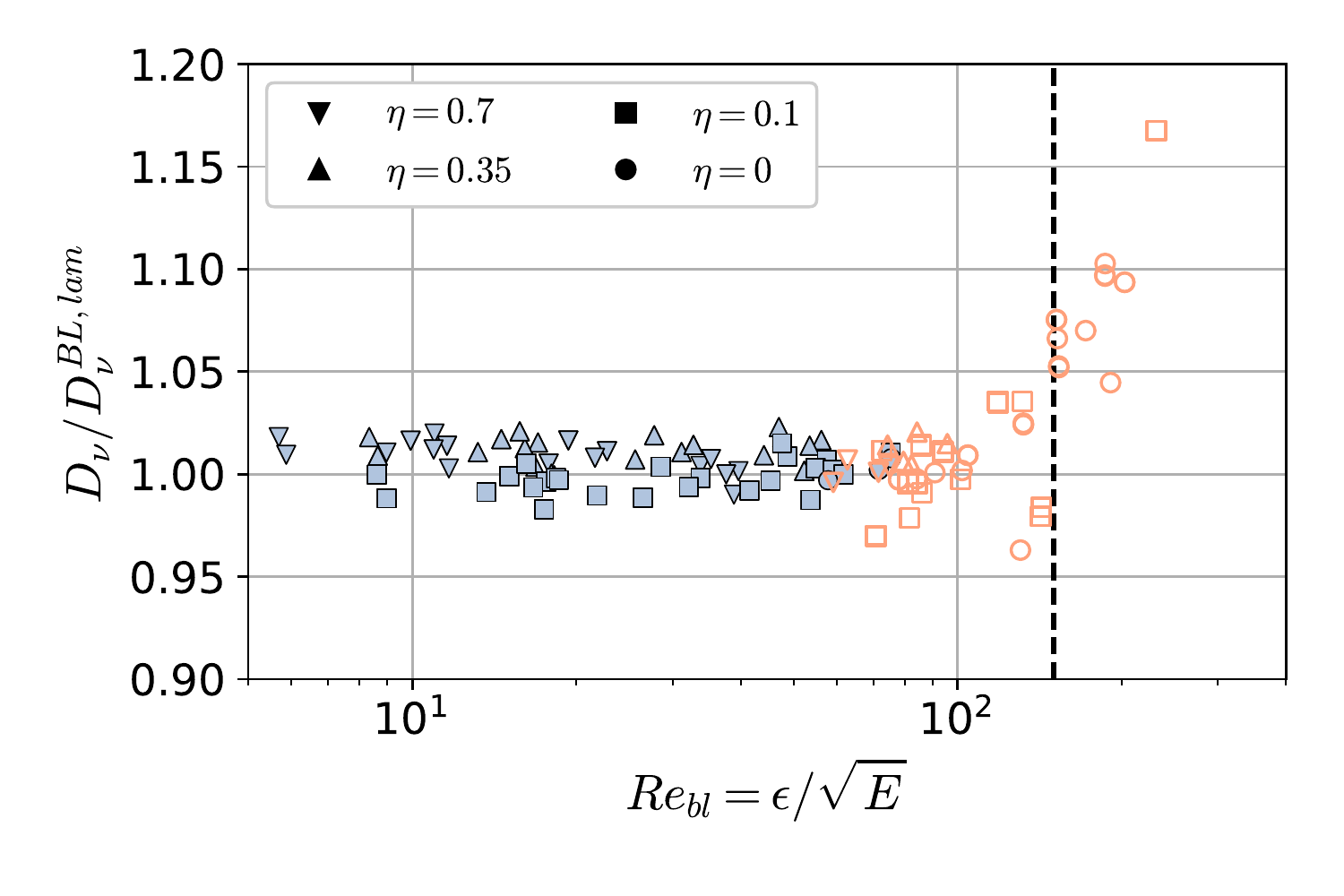}
	\caption{Total viscous dissipation $D_\nu$ in simulations normalized by the laminar dissipation $D_\nu^{BL,lam}$ computed from equation (\ref{eq:Dnusimple}) for the 172 simulations with $Po<0.05$ and $E\leq10^{-4}$ and low magnetic energy.
	Filled and open symbols represent stable and unstable flows, respectively.
	For $Re_{bl} \gtrsim 100$ we see the transition to the turbulent regime characterized by a larger dissipation.}
	\label{fig:Dnu_turb}
\end{center}
\end{figure}

More recently, the turbulent boundary-layer dissipation has been investigated by \cite{sous2013friction} showing that above a critical value of the boundary layer Reynolds number $Re_{bl}=\epsilon E^{-1/2}$ of order 150, the dissipation becomes significantly larger than the laminar one.
Concentrating on simulations with $E\leq 10^{-4}$ and $Po<0.05$, figure \ref{fig:Dnu_turb} shows the ratio of the total measured dissipation $D_\nu$ to laminar dissipation $D_\nu^{BL,lam}$ given by expression \ref{eq:Dnusimple}, as a function of the local Reynolds number.
Stable cases have their viscous dissipation accurately predicted by the laminar boundary friction.
However, shortly after the onset of the instability, for $Re_{bl}\gtrsim 100$ we observe an increase of the dissipation of up to $20\%$ compared to the laminar model, in qualitative agreement with \citet{sous2013friction}, albeit an order of magnitude smaller.
To quantitatively test their dissipation law in our setup, one should further reduce the Ekman number while giving particular attention to the measure of the uniform vorticity flow that enters the calculation of $D_\nu^{BL,lam}$.
However, reaching $Re_{bl}>400$ will be numerically challenging.

\section{Precession driven dynamos}
\label{dynamo}

\begin{figure}                
  \begin{center}
      \includegraphics[width=0.8 \linewidth]{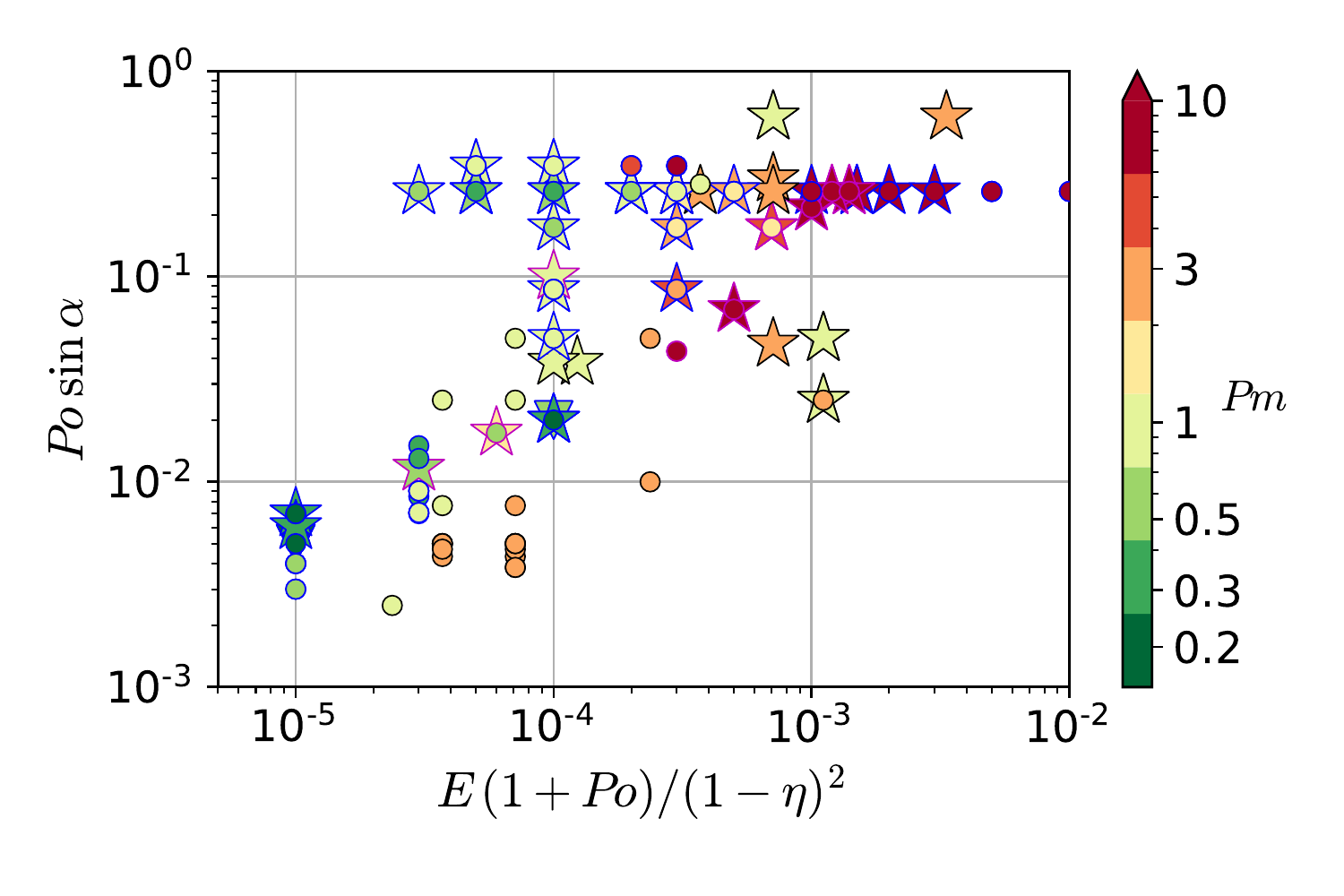}
      \includegraphics[width=0.8 \linewidth]{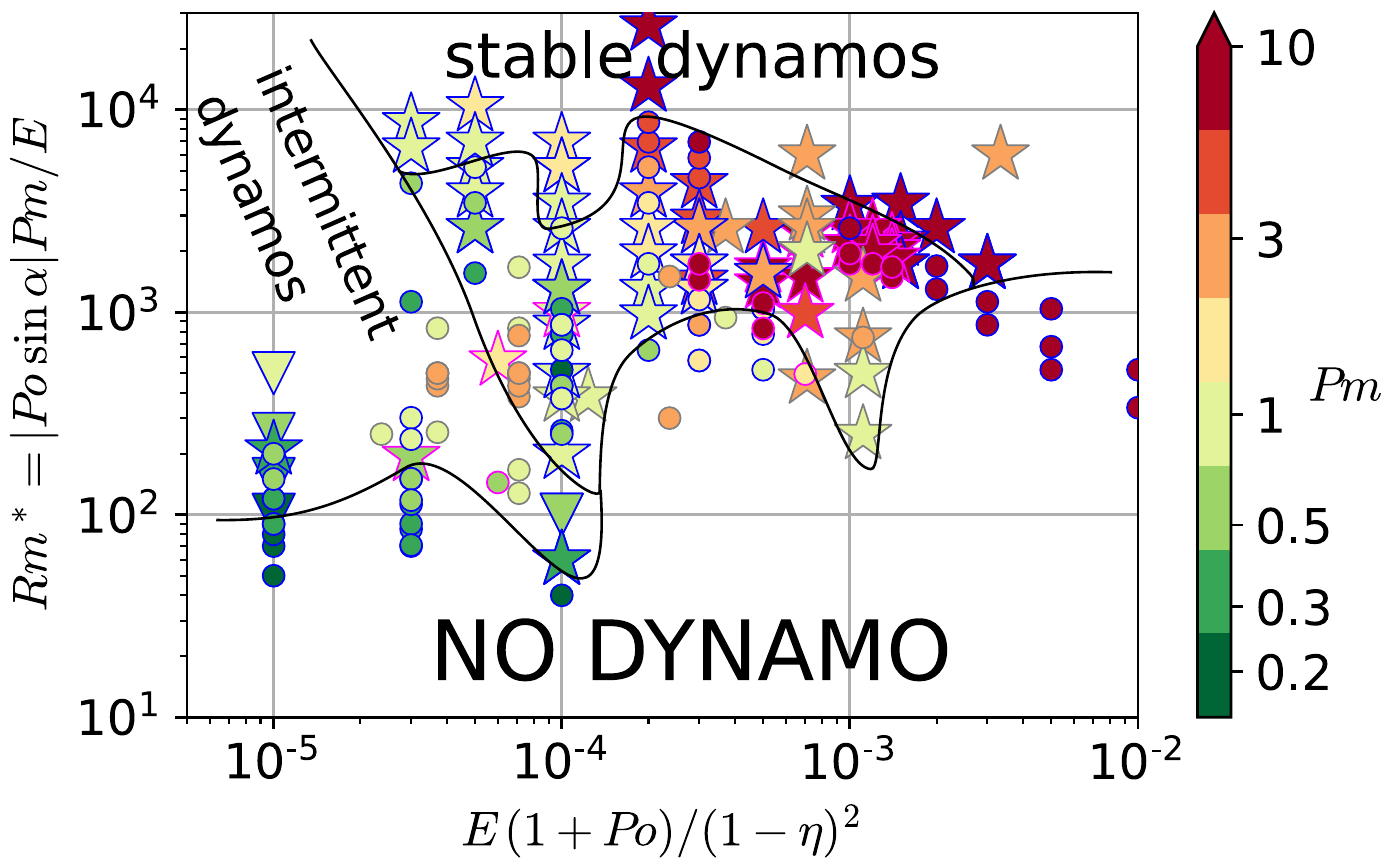}
    \caption{Saturated dynamos (stars), self-killing dynamos (triangles) and non-dynamo (circles) represented in two parameter-space planes. Symbols outlined in pink are from \cite{lin2016precession}; a grey outline indicates a conducting inner-core; a blue outline stands for a small insulating inner-core ($\eta=0.1$) or no inner-core (full sphere). The x-axis is the standard Ekman number based on the gap width.}
         \label{fig:run_summary}                
  \end{center}
\end{figure}

We now add a magnetic field and solve the whole system of equations (\ref{NS})-(\ref{eq:divB}).
In all our simulations, the initial magnetic and velocity fields are random.
We have produced two databases totaling more than 900 simulations.
The first set of more than 750 simulations consists in a broad search in the 4-dimensional parameter space ($E$, $\eta$, $Po$, $Pm$) which resulted in only few dynamos.
In this set, when an inner-core is present it is conducting with the same conductivity as the fluid. The non-dynamo runs of this set were extensively used in the previous sections.
The second set of runs uses an insulating inner-core and is focused on several 1-dimensional paths across the parameter space, yielding a few tens of self-sustained dynamos.
The explored parameters are summarized in figure \ref{fig:run_summary}, and the full databases are made freely available at \url{https://doi.org/10.6084/m9.figshare.7017137}.

\subsection{Beyond Tilgner (2005)}

\begin{figure}                
  \begin{center}
      \includegraphics[width=0.7\linewidth]{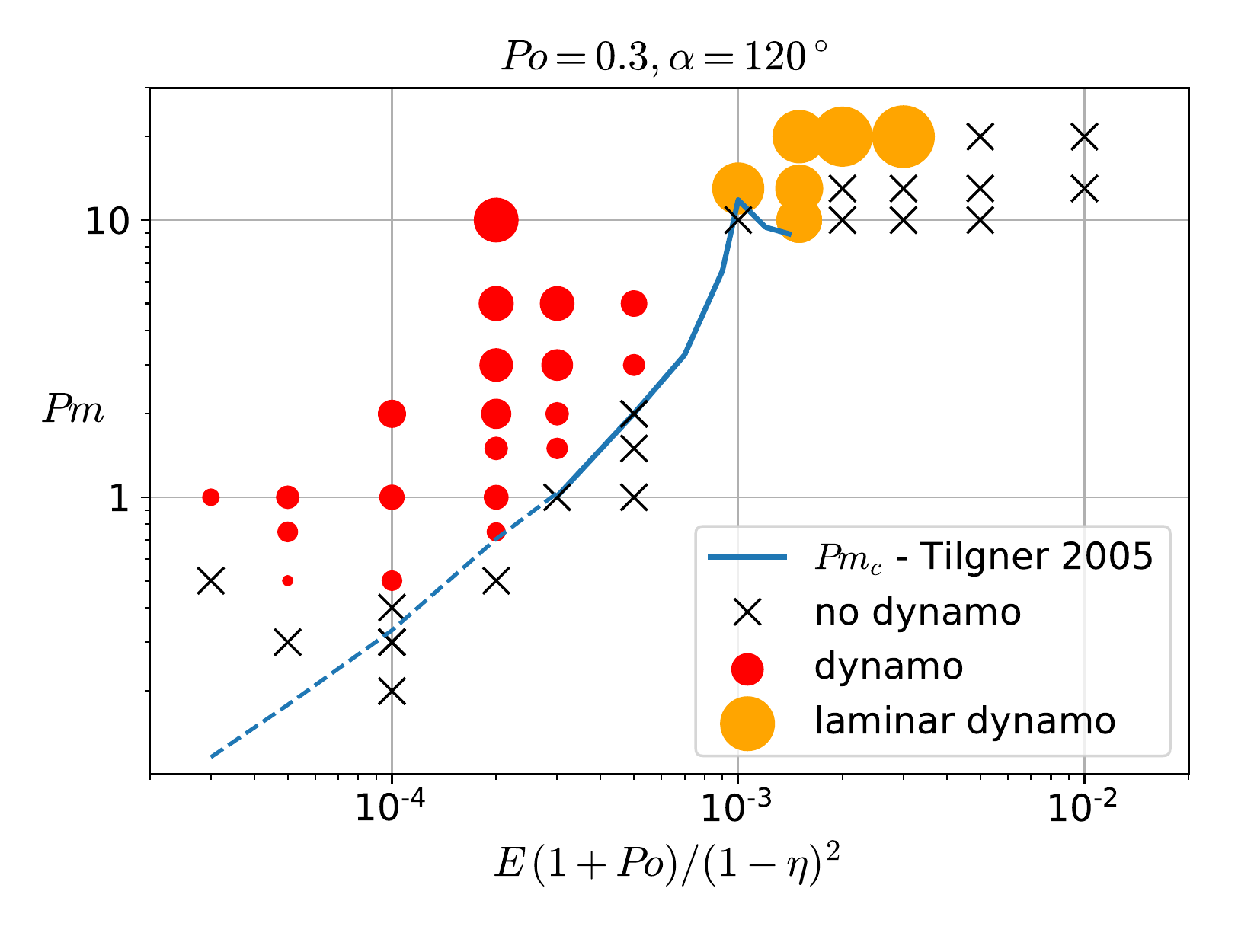}
    \caption{Successful dynamo (circle) and non-dynamo (cross) simulations for the precession parameters of \cite{tilgner2005precession}: $\eta=0.1$, $Po=0.3$, $\alpha=120^{\circ}$, with a stress-free insulating inner core. The solid blue line is the critical $Pm$ for dynamo action found by \cite{tilgner2005precession}, and the dashed blue line is the law $Pm_c = 300 E_a^{-1/2} E\,(1+Po)/(1-\eta)^2$ that he proposed (cast to our definition of $E$).
The area of the dots is proportional to the magnetic energy.}
         \label{fig:onset2}
  \end{center}
\end{figure}

\begin{figure}
  \begin{center}
      \includegraphics[width=0.4\linewidth]{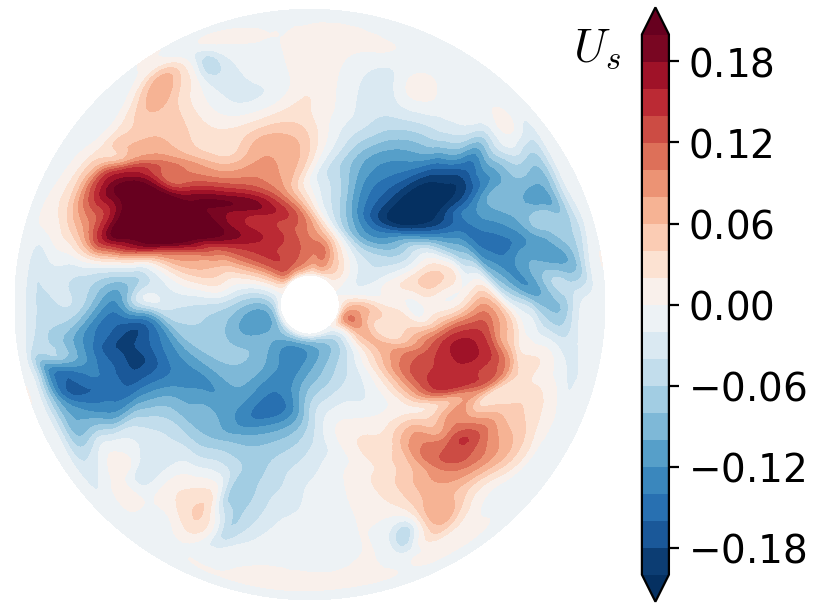}
      \includegraphics[width=0.4\linewidth]{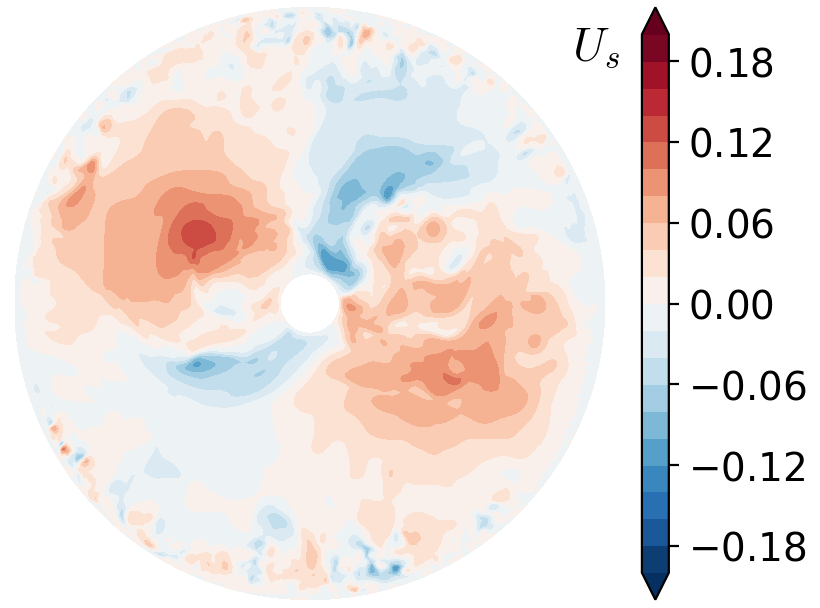}
      \includegraphics[width=0.4\linewidth]{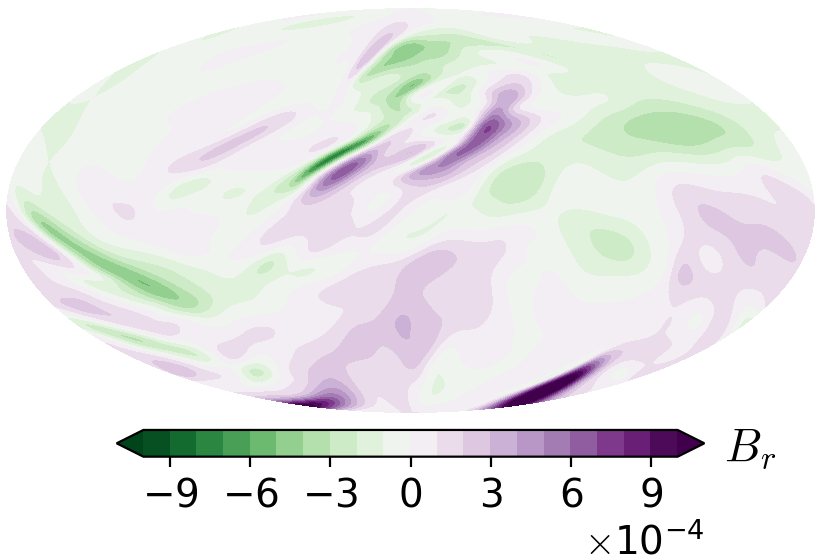}
      \includegraphics[width=0.4\linewidth]{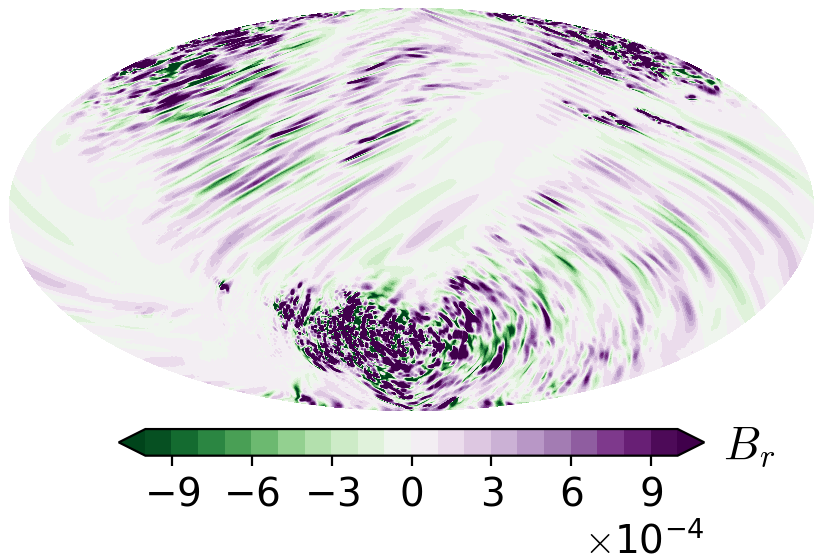}
    \caption{Radial velocity in the equatorial plane (top) and radial magnetic field at the core surface (bottom) for two different values of the viscosity, near the onset of dynamo action.
    Lower viscosity (right: $E=1.87\times 10^{-5}$, $Pm=1$) results in much smaller scales in both velocity and magnetic fields than the larger viscosity (left: $E=1.25\times 10^{-4}$, $Pm=0.75$).
    Both cases have  $Po=0.3$, $\alpha=120^\circ$ and a small insulating inner-core ($\eta=0.1$)}
         \label{fig:UB_til}
  \end{center}
\end{figure}

\begin{figure}[t]
  \begin{center}
      \includegraphics[width=0.8\linewidth]{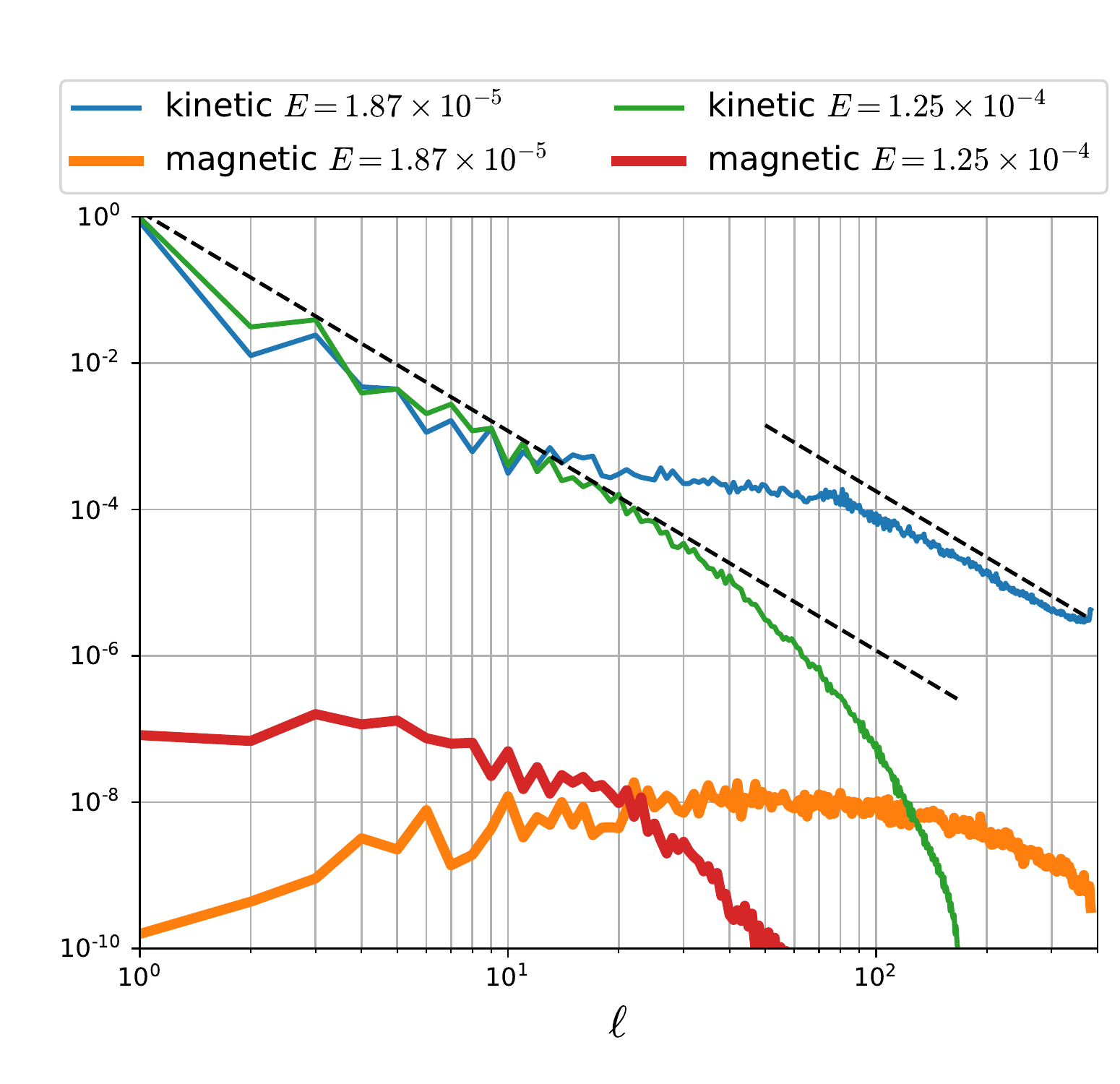}
    \caption{Magnetic and kinetic energy spectra as a function of spherical harmonic degree $\ell$ at the fluid surface for the magnetic field and below the Ekman layer for the velocity field. The black dashed lines indicate slopes of $-3$ for the kinetic energy spectra. The two cases have $Po=0.3$, $\alpha=120^\circ$, a small insulating inner-core ($\eta=0.1$) and $Pm=1$ and differ by their Ekman number $E=1.25\times10^{-4}$ and $E=1.87\times10^{-5}$, the latter displaying more energy at smaller scales.}
         \label{fig:spec_til}                
  \end{center}
\end{figure}

\cite{tilgner2005precession} has shown that precessing spheres can generate dynamos, either driven by the Ekman pumping of the forced basic laminar flow (at relatively large values of $E$), or driven by the anti-symmetric flow associated with instabilities (at smaller $E$).
More recently, \cite{lin2016precession} found that large scale vortices are sometimes generated by these instabilities, and that they contribute to magnetic field generation.

We first focus on the precession rate $Po=0.3$ and precession angle $\alpha=120^\circ$ for which the onset for dynamo action has been determined by \cite{tilgner2005precession} in his figure 4.
Our results are summarized in figure \ref{fig:onset2}, showing that our non-linear dynamos are correctly separated by his critical magnetic Prandtl number $Pm_c$ curve (the solid line in figure \ref{fig:onset2}).
Thanks to today's computing facilities and to the highly efficient XSHELLS code, we were able to further decrease the viscosity by a factor 10.
Since the forcing ($Po=0.3$) is kept constant here, this leads to turbulent flows, which require high resolutions and a high degree of parallelization to simulate.
At large and small Ekman numbers, we observe the apparition of two local minima for the critical magnetic Prandtl number $Pm_c$ of the dynamo onset: one for $E \simeq 1.5 \times 10^{-3}$ and one for $E \simeq 10^{-4}$.
If the former can be explained by the transition from base flow driven dynamos to instability driven dynamos, the latter is more difficult to interpret. Indeed, \cite{tilgner2005precession} explains the decrease of the critical magnetic Prandtl number $Pm_c$  in the range $10^{-4} \le E \le10^{-3}$ by assuming that dynamo action takes place above a critical magnetic Reynolds number $Rm_c \propto E_a^{1/2} Pm_c/E$ based on the anti-symmetric energy $E_a$.
For $E \leq 10^{-4}$, this law is no longer valid.
The turbulent fluctuations seem to have a negative effect on dynamo action, both in terms of onset ($Pm_c$ increases when decreasing $E$) and field intensity (as shown by the diminishing circle area in figure \ref{fig:onset2}).
What happens when the viscosity is lowered is illustrated in figures \ref{fig:UB_til} and \ref{fig:spec_til}.
Although the large scale flow presents similar strength and shape at $E=1.25\times10^{-4}$ and at $E=1.87\times10^{-5}$, small-scale instabilities develop near the outer shell in the latter case.
This results in a shredding of the surface magnetic field to small scales.
In figure \ref{fig:spec_til}, comparing $E=1.87\times 10^{-5}$ to $E=1.25\times 10^{-4}$, the large scale magnetic field is reduced by a factor 100, while the magnetic energy peaks at scales that will be strongly attenuated with distance, and likely to be undetectable at the planet's surface.

While small-scale turbulent fluctuations might in certain cases sustain a large-scale magnetic field \citep[by a so-called mean-field dynamo, e.g.][]{moffatt1970}, our results suggest the opposite in a precessing sphere.
In addition, a destructive effect of small-scale fluctuations on the dynamo has also been reported for precessing cubes \citep{goepfert2018mechanisms} and cylinders \citep{nore2014pamir}.
However, a constructive effect of small-scale fluctuations cannot be excluded in the very low $Pm$ regime relevant for planetary cores but out of reach with direct numerical simulations.
Further dedicated studies are needed to address this issue.

However, this path with fixed $Po=0.3$ is not appropriate for typical planetary regimes, for which $Po \ll 1$, and we now vary the precession rate.

\subsection{Varying the precession rate}

Surprisingly, when setting $Po=0.2$ or $Po=0.4$ instead of $Po=0.3$, the critical magnetic Prandtl number $Pm_c$ required for dynamo action increases (see Fig.~\ref{fig:run_summary}a).
This means that the flow generated by a precession rate $Po=0.2$ or $Po=0.4$ is less efficient than the one obtained at precession rate $Po=0.3$.
This already highlights the complicated landscape in which we are trying to find dynamos.
Specific values of the precession rate will lead to dynamos whereas neighboring values will not.
This is apparent in figure \ref{fig:run_summary}, where circles (decaying magnetic field) and stars (dynamos) are entangled.
We have not been able to find simple parameter combinations that allowed to disentangle them.
For instance, we introduce a precession-based magnetic Reynolds number $Rm^* = |Po \sin \alpha| Pm / E$.
Figure \ref{fig:run_summary}b shows that a stable dynamo is found for a low $Rm^* = 60$ (at $E=10^{-4}$, $Po=0.02$, $Pm=0.3$), while several cases at $Rm^* \ge 4000$ do not produce a magnetic field.
Furthermore, the power-based scaling laws that govern convective dynamos \citep[e.g][]{christensen2009energy,oruba2014predictive} do not work here.
A possible reason being that the power is injected by viscous coupling and that laminar viscous dissipation at the boundaries remains dominant in the accessible parameter range (see Fig. \ref{fig:Dnu_turb}).

\subsection{Low viscosity dynamos and large-scale vortices}

\begin{figure}[t]
\begin{center}
\includegraphics[width=0.8\linewidth]{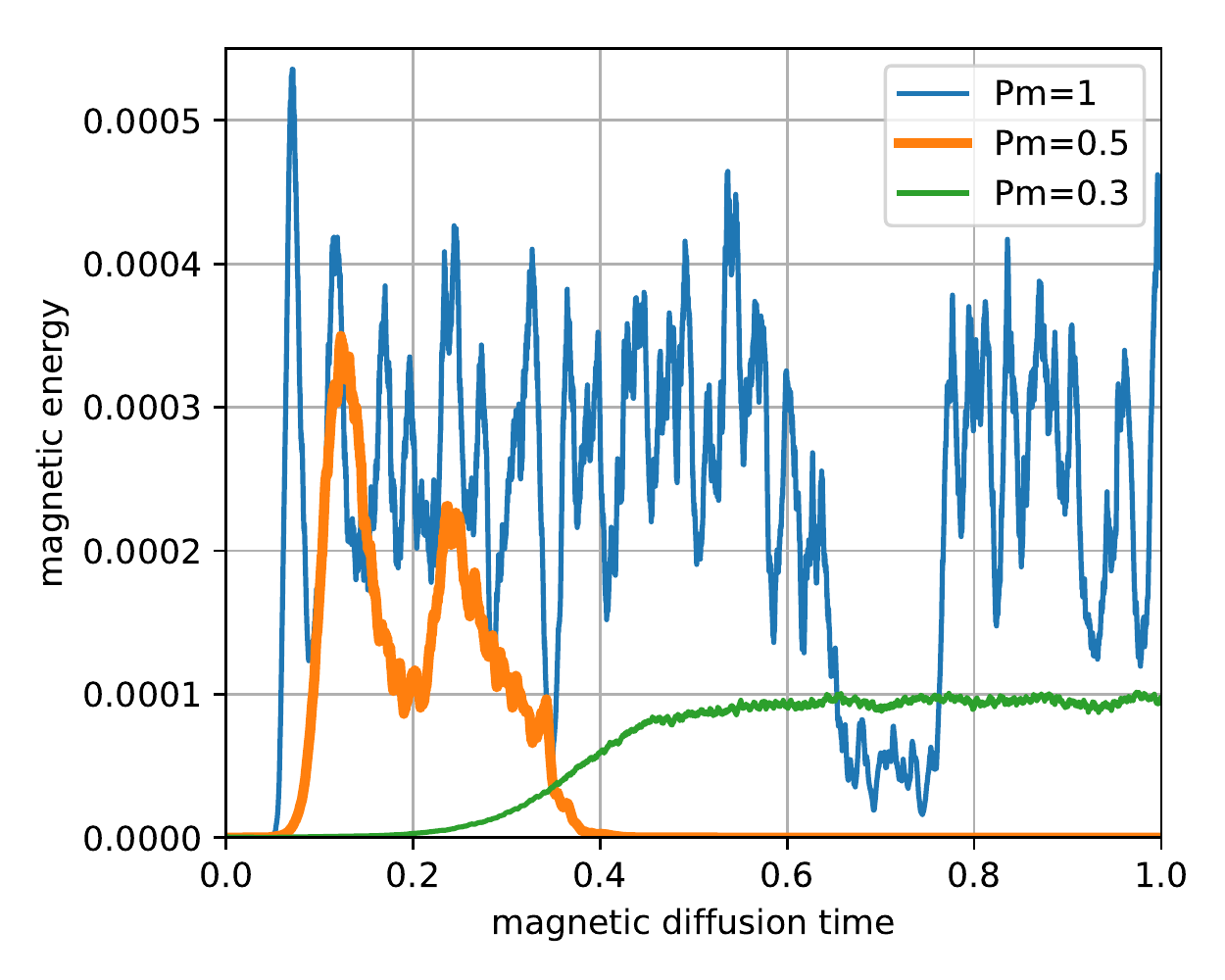}
\end{center}
\caption{Time-evolution of the magnetic energy in dynamos at $E=7.94 \times 10^{-5}$, $Po=0.02$ and $\alpha=90^\circ$, with a small insulating inner-core $\eta=0.1$. While $Pm=0.3$ is a stationary dynamo, the stronger Lorentz force at $Pm=0.5$ leads to the loss of the magnetic field. One magnetic diffusion time is $R^2(1-\eta)^2\mu\gamma$.
The link with large-scale cyclones is shown in supplementary animations \url{https://doi.org/10.6084/m9.figshare.7063652}.}
\label{fig:self-kill}
\end{figure}

\subsubsection{Stable dynamos}

Decreasing $Po$ and $E$ together, we find a few stable dynamos in full spheres and spherical shells with a small inner-core ($\eta = 0.1$).
The dynamos obtained at the lowest viscosities and forcing (low $Pm$, $E$, $Po$) are all associated with large-scale vortices (LSV, see figure \ref{fig:LSV001} and \ref{fig:lsv} for examples).
The importance of LSV for dynamo action has been already highlighted by \cite{lin2016precession}, and our study confirms that they play an important role for dynamo action \citep[see also][in the context of rotating convection]{guervilly2015generation}.
With stable, persistent LSV, the magnetic energy is rather stable (case $Pm=0.3$ in figure \ref{fig:self-kill}), allowing to obtain dynamos at low viscosity ($Pm < 1$, $E \leq 10^{-4}$), seemingly relevant for planetary cores.

As an example, at $E=7.94\times 10^{-5}$, $Po=0.02$, $\alpha=90^\circ$, we found a stable saturated dynamo at $Pm=0.3$ (see figure \ref{fig:self-kill}).
Three stable LSV are seen during the whole simulation, unaffected by the Lorentz force.
Furthermore, the fluid rotation vector does not change significantly from the corresponding hydrodynamic case.

\subsubsection{Self-killing dynamos}

\begin{figure}
\begin{center}
\includegraphics[width=0.8\linewidth]{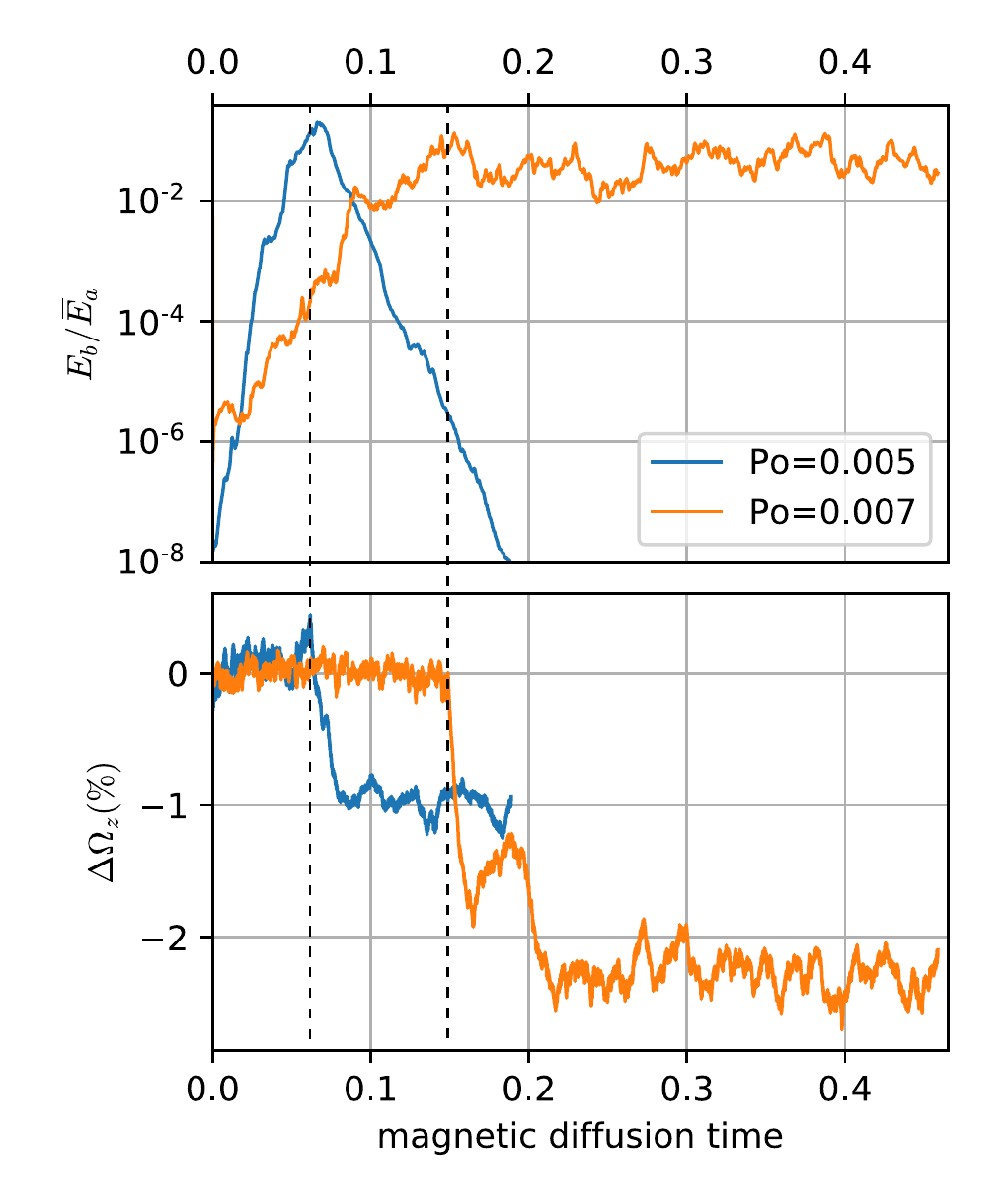}
\end{center}
\caption{Top: ratio of magnetic energy $E_b$ over time-averaged antisymmetric kinetic energy $\bar{E_a}$. Bottom: relative variations (in percent) of the projection of the fluid rotation axis on the planet spin axis. Both cases have $E=10^{-5}$, $\alpha=90^\circ$, $Pm=0.3$ and no solid inner-core. They differ only by their precession rate $Po$.
The saturation of the magnetic energy at $E_b \approx 0.1\bar{E_a}$ leads to a change in the rotation axis of the fluid, as hinted by the vertical dashed lines.
In the case $Po=0.005$, this change seems permanent despite the loss of the magnetic field.
The link with large-scale cyclones is shown in supplementary animations \url{https://doi.org/10.6084/m9.figshare.7063652}.}
\label{fig:eb_vs_wz}
\end{figure}

However, when increasing the electrical conductivity to $Pm=0.5$, after the exponential growth of the magnetic field, the Lorentz force becomes strong enough to alter the flow so that the magnetic field decays and never recovers, even after the magnetic field has decayed to very low intensity (case $Pm=0.5$ in figure \ref{fig:self-kill}).
While the three LSV are present in the growing phase, they wither away when the magnetic field reaches saturation value.

We found other self-killing precessing dynamos as indicated by the triangles in figure \ref{fig:run_summary}.
At $E=10^{-5}$, $Po=0.005$, $\alpha=90^\circ$, $\eta=0$, large-scale vortices are observed together with a growing magnetic field for $0.2 \le Pm \le 1$ until the Lorentz force kills the vortices and the magnetic field immediately decays.
This is illustrated in figure \ref{fig:eb_vs_wz}a.
For the limited time we could run these self-killing dynamos, the LSV and hence also the magnetic field have not been able to recover, even though the field has reached levels where the Lorentz force is negligible.
Figure \ref{fig:eb_vs_wz}b shows that the mean rotation axis of the fluid changes slightly but permanently when the magnetic field reaches its peak intensity.
We hypothesize that the slight change in rotational state of the fluid is enough to prevent the reformation of the LSV.
Furthermore, the system stays around the second rotational state even though the magnetic field has vanished, which suggests a hydrodynamic bistability.

Self-killing dynamos have already been reported in simple laminar dynamo models \citep{fuchs1999} or turbulent experiments \citep{miralles2015}.
To our knowledge, it is the first time such self-killing dynamos are reported in a self-consistent, turbulent setup.
Hence, it highlights that kinematic precession dynamos (a growing magnetic field without the Lorentz force) do not imply that strong magnetic fields can be sustained once the Lorentz force is taken into account.

\subsubsection{Intermittent dynamos}

\begin{figure}
\begin{center}
\includegraphics[width=0.75\linewidth]{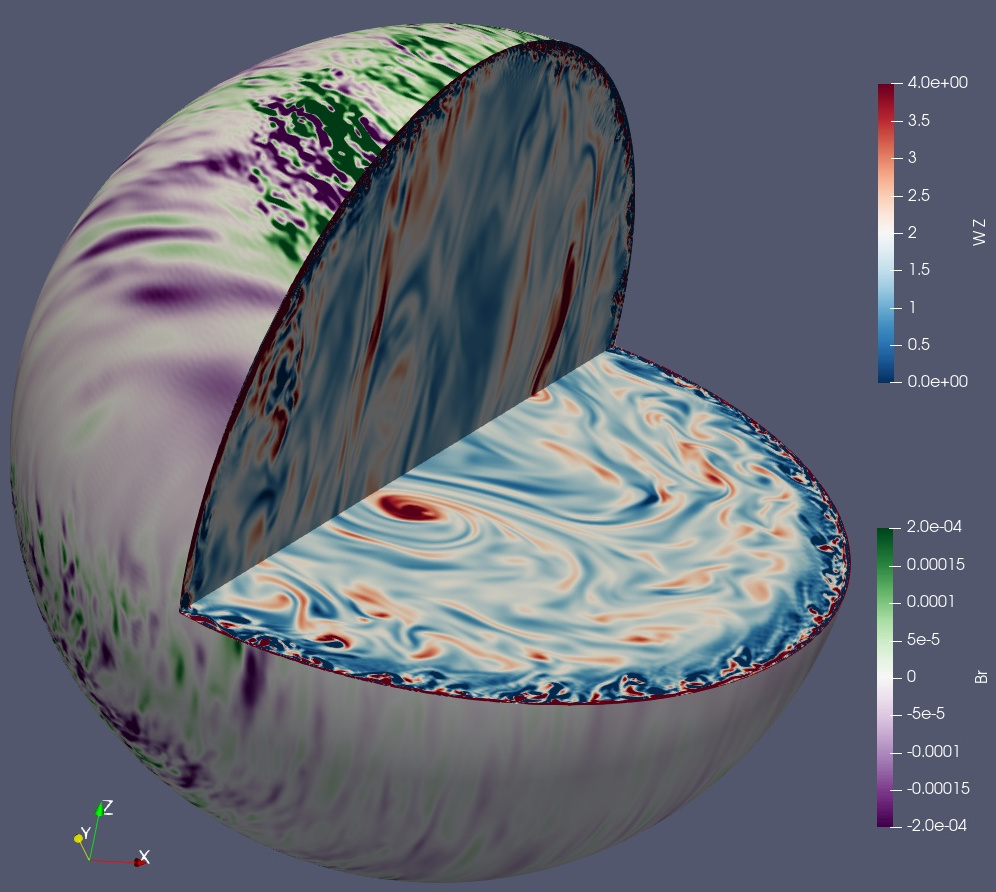}
\includegraphics[width=0.75\linewidth]{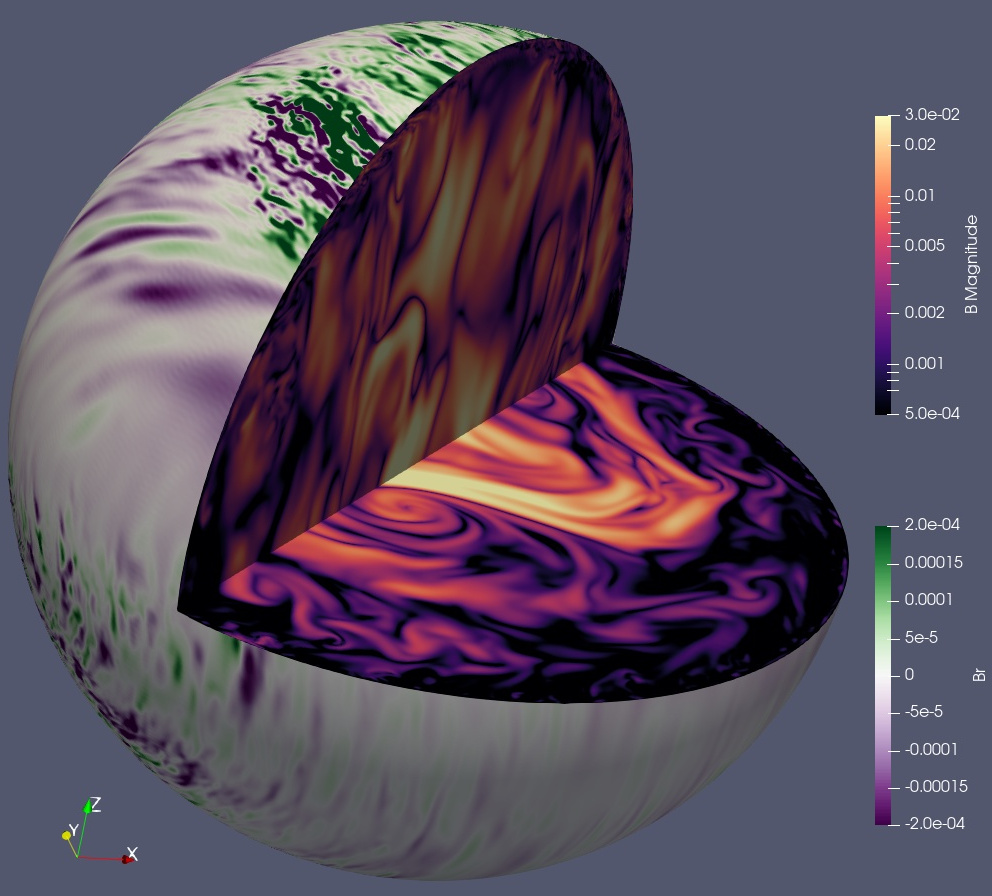}
\end{center}
\caption{Snapshots (corotating with the fluid) at $E=10^{-5}$, $Pm=0.3$, $Po=0.007$, $\alpha=90^\circ$ and no solid inner-core. The radial magnetic field is shown on the surface. In the bulk: the axial vorticity along fluid rotation axis (top) and the magnetic intensity (bottom -- in logarithmic scale).}
\label{fig:trophy}
\end{figure}

In addition, many other dynamos show large fluctuations of their magnetic energy of about a factor 10 to 100, suggesting that the Lorentz force is often pushing the flow to a different attractor before quickly recovering.
When the magnetic energy is low, the LSV develop and the magnetic field can grow.
When the magnetic energy saturates at a high enough level, the Lorentz force sometimes kills the LSV and thus the magnetic energy decays to a lower level.
This behavior is rather common and illustrated by case $Pm=1$ in figure \ref{fig:self-kill}, and case $Po=0.007$ in figure \ref{fig:eb_vs_wz}.
At the planet's surface, this may appear as an intermittent dynamo, with stronger magnetic field alternating with undetectable magnetic field.

\subsubsection{Small-scale surface magnetic field at low viscosity}

\begin{figure}
\begin{center}
\includegraphics[width=0.7\linewidth]{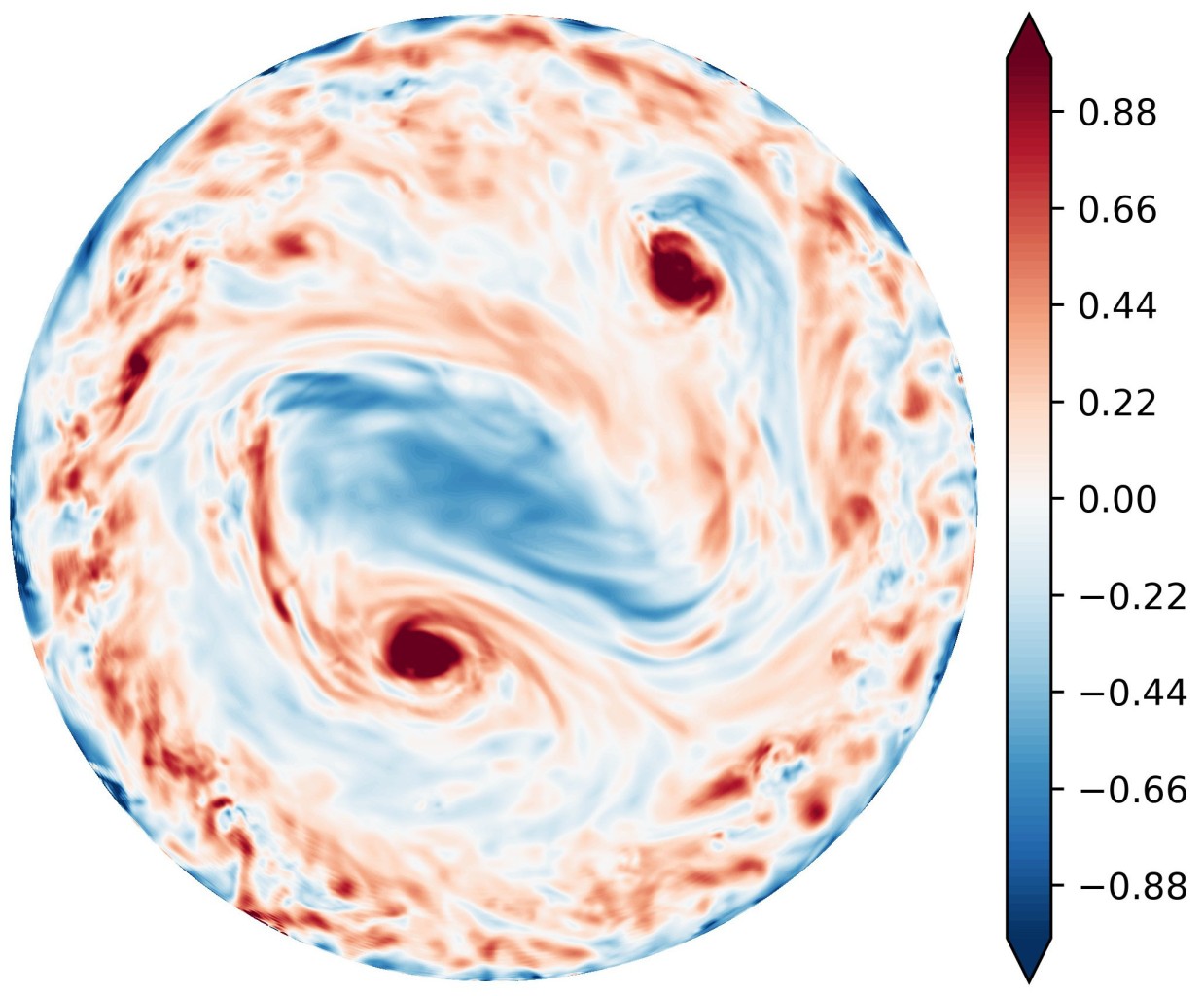}
\end{center}
\caption{Vorticity along the fluid rotation axis in the frame rotating with the fluid, averaged along the fluid rotation axis, for $E=10^{-5}$, $Po=0.007$, $\alpha=90^\circ$, $Pm=0.3$. Turbulent boundary layers have been excluded from the average. Two intense cyclones (red) can be seen around a large central anti-cyclonic (blue) region.}
\label{fig:lsv}
\end{figure}

Keeping $E=10^{-5}$, $\alpha=90^\circ$, $\eta=0$ and increasing the precession rate from $Po=0.005$ -- a self-killing dynamo -- to $Po=0.007$, the LSV are now able to withstand the Lorentz force, and at $Pm=0.3$ the dynamo saturates with the magnetic energy fluctuating within a factor 10.
This is the stable dynamo we obtained with parameters closest to planetary values, and we double-checked by also computing it in the mantle frame with lower time steps (see \S\ref{sec:num}).
Time-evolution of the magnetic energy is shown in figure \ref{fig:eb_vs_wz}a.
A snapshot of the corresponding flow and field is shown in figure \ref{fig:trophy}, while the LSV are highlighted in figure \ref{fig:lsv}.
Two large-scale cyclones are seen in the bulk, together with small-scale vorticity fluctuations.
Near the outer shell, a thick layer, much thicker than the Ekman layer, of intense small-scale vorticity is seen.
This leads to a small-scale magnetic field at the surface, while the larger-scale, stronger field does not escape the bulk. This shift towards small-scale surface magnetic field seems robust as the viscosity is decreased toward planetary values.

Note however that, at even lower $Pm$, we cannot exclude the emergence of a large-scale magnetic field produced by the small-scale turbulence \citep[see e.g.][]{moffatt1970}.

\section{Application to the Moon}\label{appliMoon}

\begin{figure}
\begin{center}
 	\begin{tabular}{cc}
\subfigure[]{
\includegraphics[width=0.7\linewidth]{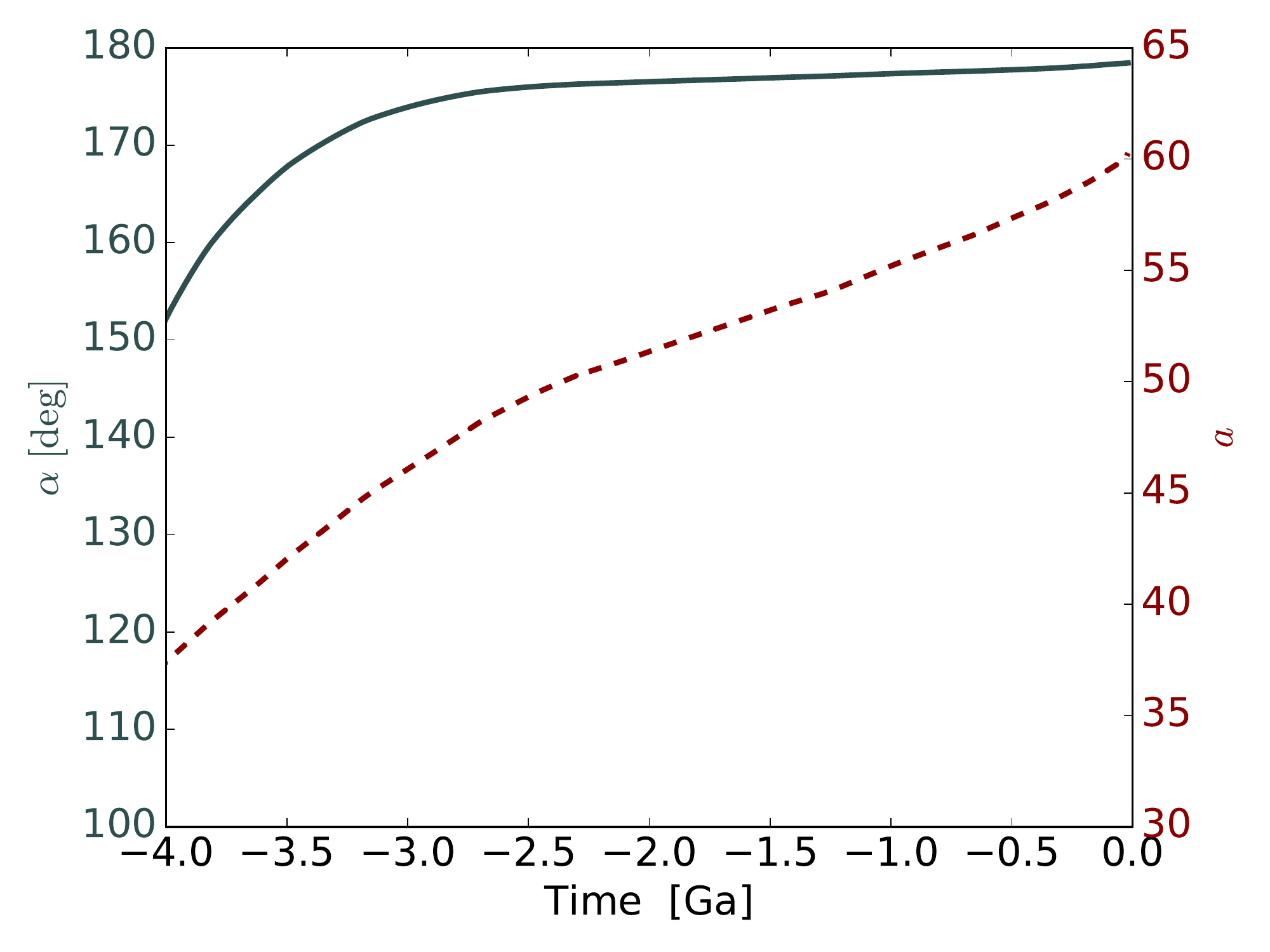}}\\
\subfigure[]{
\includegraphics[width=0.7\linewidth]{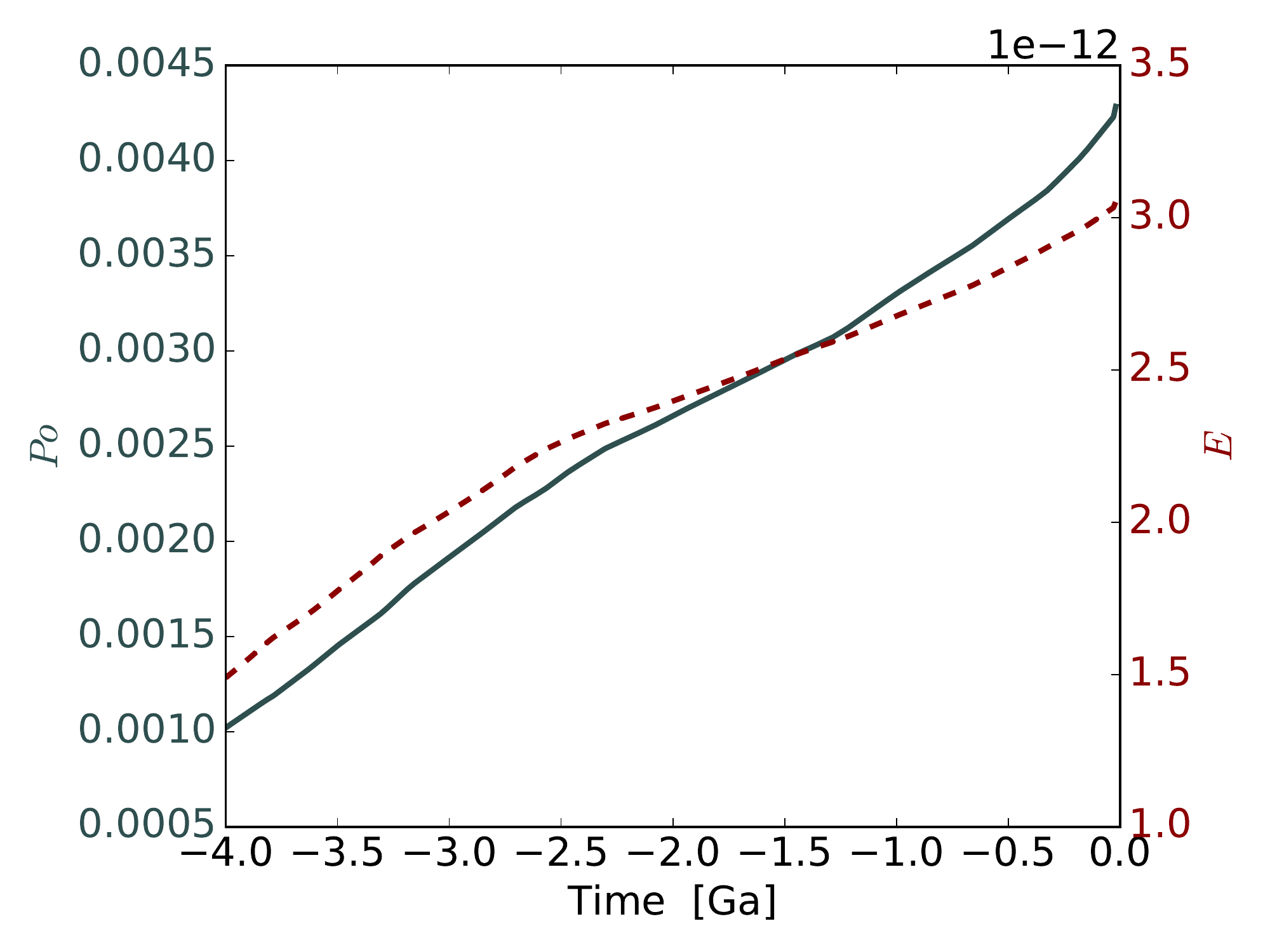} } 
\end{tabular}
\caption{Time evolution of lunar values. (a) $\alpha$ (solid line), $a$ (dashed line).  Note that \cite{dwyer2011long} consider negative values of $Po$ with $\alpha \in [0,90^{\circ}]$ whereas, here, our convention is to consider positive $Po$ and $\alpha>90^{\circ}$ for retrograde precession. (b) $Po$ (solid line) and $E$ (dashed line), where $E$ is calculated for the liquid core (with $R=350\, \mathrm{km}$, $\nu=10^{-6}\, \mathrm{m}^2.\mathrm{s}^{-1}$).}
\label{fig:Moon}
\end{center}
\end{figure}

\subsection{Time evolution of the lunar precession}

Precession has been suggested for driving turbulence and dynamo magnetic fields in the past Moon liquid core \citep{dwyer2011long}.
In the light of the present results, we propose to revisit the time evolution of precession driven flows in the lunar core.
In the following, we will consider a lunar metallic liquid core of radius $R=350\, \mathrm{km}$.

During the Moon history, the variation of the precession angle $\alpha$ can be related to the variation of the semi-major axis $a$ of the lunar orbit by \cite[Eq. 5 of][]{dwyer2011long}
\begin{eqnarray}
180^{\circ}-\alpha &=& 0.1075^{\circ} \tilde{a}^{10}-0.0332^{\circ} \tilde{a}^9-1.0008^{\circ} \tilde{a}^8+0.6110^{\circ} \tilde{a}^7+2.7016^{\circ} \tilde{a}^6-1.7281^{\circ} \tilde{a}^5 \nonumber \\
& & -2.3280^{\circ} \tilde{a}^4-1.4509^{\circ} \tilde{a}^3+6.9951^{\circ} \tilde{a}^2-6.6208^{\circ} \tilde{a}+5.5828^{\circ} ,
\end{eqnarray}
where $\alpha$ is given in degrees, $\tilde{a}=(a/R_E-46.6308)/7.7288$ with the Earth radius $R_E$ (this formula gives $\alpha$ for  $34.2 \leq a/R_E \leq 60.2$). Then, $a$ can be related to time using the so-called nominal model of \cite{dwyer2011long}, shown in their figure S2 and reproduced in Fig \ref{fig:Moon}(a). 

Assuming that the Moon remains synchronized during its history, the lunar spin rate $\Omega_s$ is given by its orbital rate. Thus, using the Kepler law $\Omega_s \propto a^{-3/2}$, one can calculate $E$. Then, we obtain $Po$ by extracting the lunar precession rate $\Omega_p$ from the Fig. 19 of \cite{touma1994evolution}. The time evolution of $Po$ and $E$ over the lunar history are presented in Fig. \ref{fig:Moon}(b).

\subsection{Flow stability at the lunar CMB during its history}

\begin{figure}
\begin{center}
\includegraphics[width=0.7\linewidth]{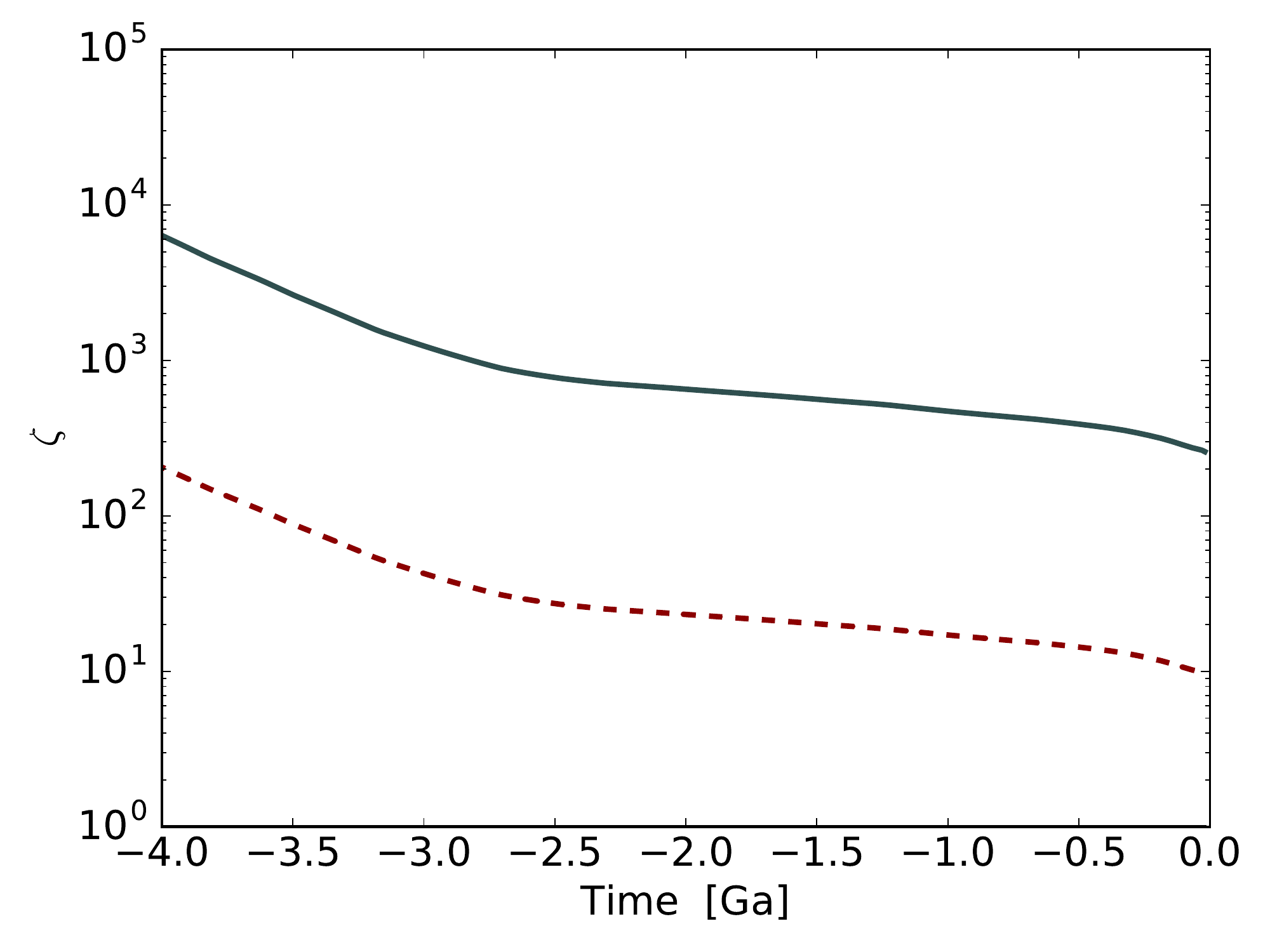} 
\caption{Stability of the lunar liquid core considering the BL-CMB (solid line), with $K_{BL}=60$, and the CSI-CMB (dashed line), with $K_{CMB}=8$. Each instability is expected for $\zeta>1$. Since $\zeta=\epsilon/(K_{BL} E^{1/2})$ for the BL-CMB, the local Reynolds number $Re=\epsilon/E^{1/2}$ at the lunar CMB is also given by the solid line, with a factor $K_{BL}=60$.}
\label{fig:stabMoon}
\end{center}
\end{figure}

One can calculate the stability of the lunar liquid core for the parametric instability (CSI-CMB) and the boundary layer instability (BL-CMB).
To do so, we define a general parameter $\zeta$ as an estimate for the onset distance, given by $\zeta=\epsilon/(K_{CMB} E^{3/10})$ for the CSI-CMB and $\zeta=\epsilon/(K_{BL} E^{1/2})$ for the BL-CMB. An instability is thus expected in both cases when $\zeta>1$. The results are shown in Fig. \ref{fig:stabMoon}. It confirms the fact that a CSI-CMB can be currently expected in the Moon, as already proposed by \cite{lin2015shear}.
Beyond this confirmation, this figure furthermore shows that both instabilities are clearly expected during the whole lunar history, with a BL-CMB significantly more unstable.

\subsection{Turbulent torque and dissipation in the lunar liquid core}
According to Fig. \ref{fig:stabMoon}, the local Reynolds number $Re=\epsilon/E^{1/2}$ at the lunar CMB has decreased during the lunar evolution, from $Re=4.10^5$, 4 Ga ago, to its current value of $Re=10^4$.
These values are well in the regime, in which \citet{sous2013friction} observe turbulent Ekman layers.
A turbulent friction can thus be expected at the lunar CMB, as previously suggested for the current lunar core \citep{williams2001lunar}. Naturally, a different friction would lead to a different differential rotation strength $\epsilon$, and we should thus calculate $\epsilon$ in a self-consistent way in presence of a modified (turbulent) friction.
To do so, we simply replace the viscous (laminar) term $\mathcal{L} \boldsymbol{\Gamma}_{\nu}$ in equations (\ref{eq_axi_P_x})-(\ref{eq_axi_P_z}) by the following turbulent damping term:
\begin{eqnarray}\label{turbvisc}
\mathcal{L} \boldsymbol{\Gamma}_{\nu}=   
\lambda_t \, ||\boldsymbol{\Omega} - \vect{\tilde{\Omega}_s} ||  \,(\boldsymbol{\Omega} - \vect{\tilde{\Omega}_s} ),
\end{eqnarray}
with a coefficient $\lambda_t$. The dimensionless dissipation can thus generally be written as \begin{eqnarray}\label{turbvisc23}
D_{\nu}=-I_c \, \lambda_t \, \epsilon^3,
\end{eqnarray}
where $\lambda_t$ remains to be obtained. Obtaining expressions of $\lambda_t$ requires to describe the (local) turbulent stress associated to the shear velocity $\boldsymbol{v_{sh}}= (\boldsymbol{\Omega} - \vect{\tilde{\Omega}_s} ) \times \boldsymbol{r}$ generated by the differential rotation $\boldsymbol{\Omega} - \vect{\tilde{\Omega}_s} $ at the inner and outer boundaries. Noting $\boldsymbol{\tau_v}$ the (local) surface stress per unit of mass, one can write $\boldsymbol{\tau_v}=-\kappa |\boldsymbol{v_{sh}}| \boldsymbol{v_{sh}}$, where $\kappa$ can be seen as a (local) drag coefficient. To close the equations, we need to specify $\kappa$, where the physics of the friction coupling is hidden. 

Focusing first on laminar flows, the model of \cite{sous2013friction} predicts such flows for $Re \lesssim 150$ and prescribes $\kappa=\xi \epsilon \sqrt{E}/v_{sh}$, where $\xi$ is a constant of order unity ($\xi=1$ in \citet{sous2013friction}). One can then calculate the associated viscous torque $\boldsymbol{\Gamma}_{\nu}=\int_S \boldsymbol{r} \times \boldsymbol{\tau_v} \mathrm{d}S$ on the surface $S$ of the fluid boundary. Noting the colatitude $\theta$, we have $ |\boldsymbol{r} \times \boldsymbol{\tau_v}| \propto r^3 \sin^3 \theta $, which gives
\begin{eqnarray}
\lambda_t = -5\, \xi\, \frac{1+ \eta^4 \vartheta}{1-\eta^5}\,  \frac{E^{1/2}}{\epsilon}, \label{eq:lfm5}
\end{eqnarray}
with $\vartheta=1$ (resp. $\vartheta=0$) for a no-slip  (resp. stress-free) inner boundary. For $Ek \ll 1$ and no-slip boundaries, our equation (\ref{eq:Dnusimple}) is exactly recovered with $\xi=2.62/5 \approx 0.52$, including the correct dependency in $\eta$ (see equation \ref{eq:dety00}). In the model of \cite{sous2013friction}, using this value of $\xi$ allows thus to switch naturally from the validated laminar dissipation (\ref{eq:Dnusimple}) to a turbulent dissipation.

In the turbulent regime, it is usually assumed that $\kappa$ does not vary in space. Under this hypothesis, the associated viscous torque is then given by
\begin{eqnarray}
\boldsymbol{\Gamma}_{\nu}=   
\frac{3 \pi^2}{4} (1+\eta^5) \kappa   |\boldsymbol{v_{sh}}| \boldsymbol{v_{sh}}, \label{turbvisc3}
\end{eqnarray}
which recovers equation (55) of  \cite{williams2001lunar}, obtained in the particular case $\eta=0$. Using equation (\ref{turbvisc}), equation (\ref{turbvisc3}) leads to
\begin{eqnarray}
\lambda_t = -\frac{45 \pi}{32 }\, \frac{1+\eta^5 \vartheta}{1-\eta^5}\, \kappa , \label{eq:lltt}
\end{eqnarray}
in the turbulent regime (with $\vartheta$ defined as above). Note the different dependency in $\eta$ compared to equation (\ref{eq:lfm5}).

\begin{figure}[t]
\begin{center}
\includegraphics[width=0.8\linewidth]{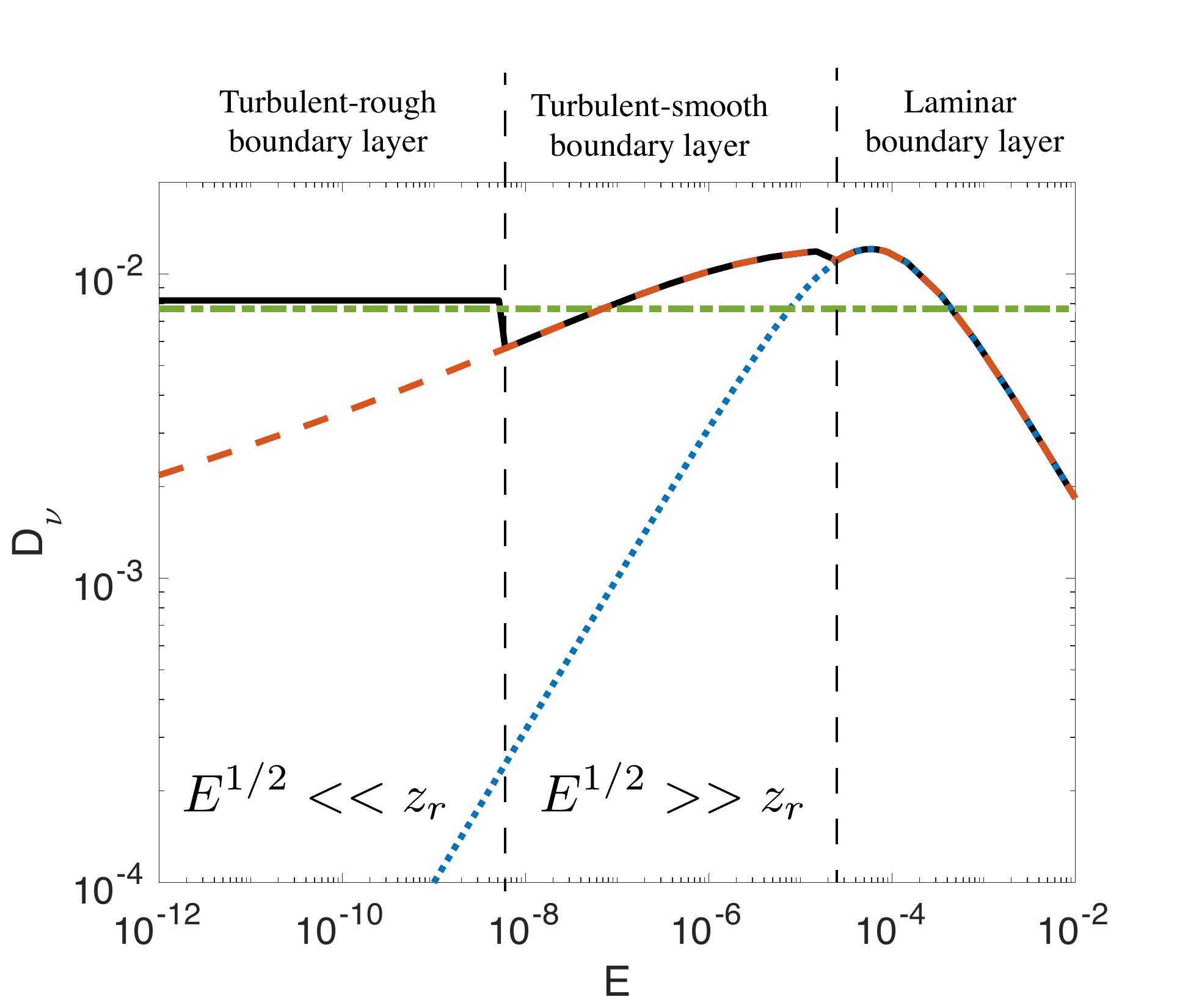}
\caption{Dimensionless dissipation of the model based on \cite{sous2013friction}, for an (arbitrary) illustrating case, in presence of a smooth CMB (dashed red line) or a rough CMB with $z_r=10^{-5}$ (solid black line), using the parameters $Po=0.02$, $\alpha=\pi/3$, $\eta=0$, $(A,B)=(3.3,3)$, $\xi=0.52$. We also show, as a horizontal dash-dotted green line, the turbulent model of \cite{yoder1981free}, and, as a dotted blue line, the laminar reduced model (\ref{eq:Dnusimple}) given by equation (\ref{eq:RotDif00}), using $\lambda_r=-2.62$, $\lambda_i=0$.}
\label{fig:MoonDissip}
\end{center}
\end{figure}

To close the equations in the turbulent regime, a value has to be chosen for $\kappa$ (assumed to be uniform in space). As a first approach, \cite{yoder1981free} simply considered a constant $\kappa=0.002$ based on \citet{bowden1953note}.
Later, refined models based on turbulent non-rotating boundary-layer theory have been proposed \citep{yoder1995venus,williams2001lunar}.
By contrast, for rotating turbulent flows ($Re>150$), \cite{sous2013friction} proposed a self-consistent approach where $\kappa= C_d \cos \alpha_0$ depends on $\epsilon$ through
\begin{eqnarray}
C_d= \frac{u_{0 \star}^2}{\epsilon^2}=\frac{k^2}{\left(\ln \left| \frac{u_{0 \star}}{2 z_0} \right|-A \right)^2+B^2}, \label{eq:sous}
\end{eqnarray}
and $\sin \alpha_0=(B u_{0 \star})/(k \epsilon)$.
Here, $k=0.4$ is the von Karman constant, $(A,B)$ are constants obtained from measurements, $u_{0 \star}$ is the unknown friction velocity and $\alpha_0$ is the so-called cross-isobar angle from the geostrophic flow due to the tilt of the velocity vector in the boundary layer.
From atmospheric measurements, we typically have $(A,B)=(1.3,4.4)$, whereas laboratory experiments rather give $(A,B)=(3.3,3)$ \citep{sous2013friction}.
Noting $z_r$ the dimensionless root mean square roughness height of the boundary, $z_0=0.11 E/u_{0 \star}$ for a smooth CMB (i.e. $z_r u_{0 \star}/E<60$), whereas $z_0=z_r/30$ for a rough CMB (i.e. $z_r u_{0 \star}/E>60$).
From equation (\ref{eq:lltt}) and (\ref{eq:sous}) we obtain $\lambda_t$, which we substitute in (\ref{turbvisc3}) to obtain the turbulent viscous torque.
We finally self-consistently solve for the torque balance (\ref{eq_axi_P_x})-(\ref{eq_axi_P_z}).
This model is implemented in the updated FLIPPER program \cite[initially introduced in][]{cebron2015bistable}, a MATLAB script calculating the theoretical uniform vorticity flow in precessing ellipsoids (\url{https://www.mathworks.com/matlabcentral/fileexchange/50612-flipper}).

Let's consider a generic case to illustrate a possible scenario for the viscous dissipation as we lower the Ekman number from numerically accessible values to planetary settings.
In this example show in Fig. \ref{fig:MoonDissip}, ($Po$, $\alpha$, $\eta$) are fixed to arbitrary values and the transition from smooth to rough boundary, i.e. $z_r u_{0\star}/E\sim 60$, is taken to arise at $E \simeq 10^{-8}$.
At moderate Ekman number $E\gtrsim 10^{-5}$ the flows remains weakly non-linear such that the dissipation is dominated by the laminar processes in the boundary layer leading to $D_{\nu}\propto I_c \epsilon^2 \sqrt{E}$.
At low enough Ekman number $E \lesssim 10^{-5}$, the boundary layer becomes turbulent for the precession parameters considered in this example, and the dissipation is then estimated using the friction model derived from \citet{sous2013friction}.
Two regimes must be distinguished, in the range $10^{-8}\lesssim E \lesssim 10^{-5}$ the roughness is buried in the boundary layer and the dissipation remains weakly dependent on the Ekman number, decreasing with $E$.
When the boundary layer thickness becomes small compared to the roughness of the boundary, $z_r$ becomes the relevant length scale for the dissipation, leading to $D_{\nu}\propto I_c \epsilon^3$, independent of E, as proposed by \citet{yoder1981free} for the lunar core.

\begin{figure}[t]
\begin{center}
\includegraphics[width=0.8\linewidth]{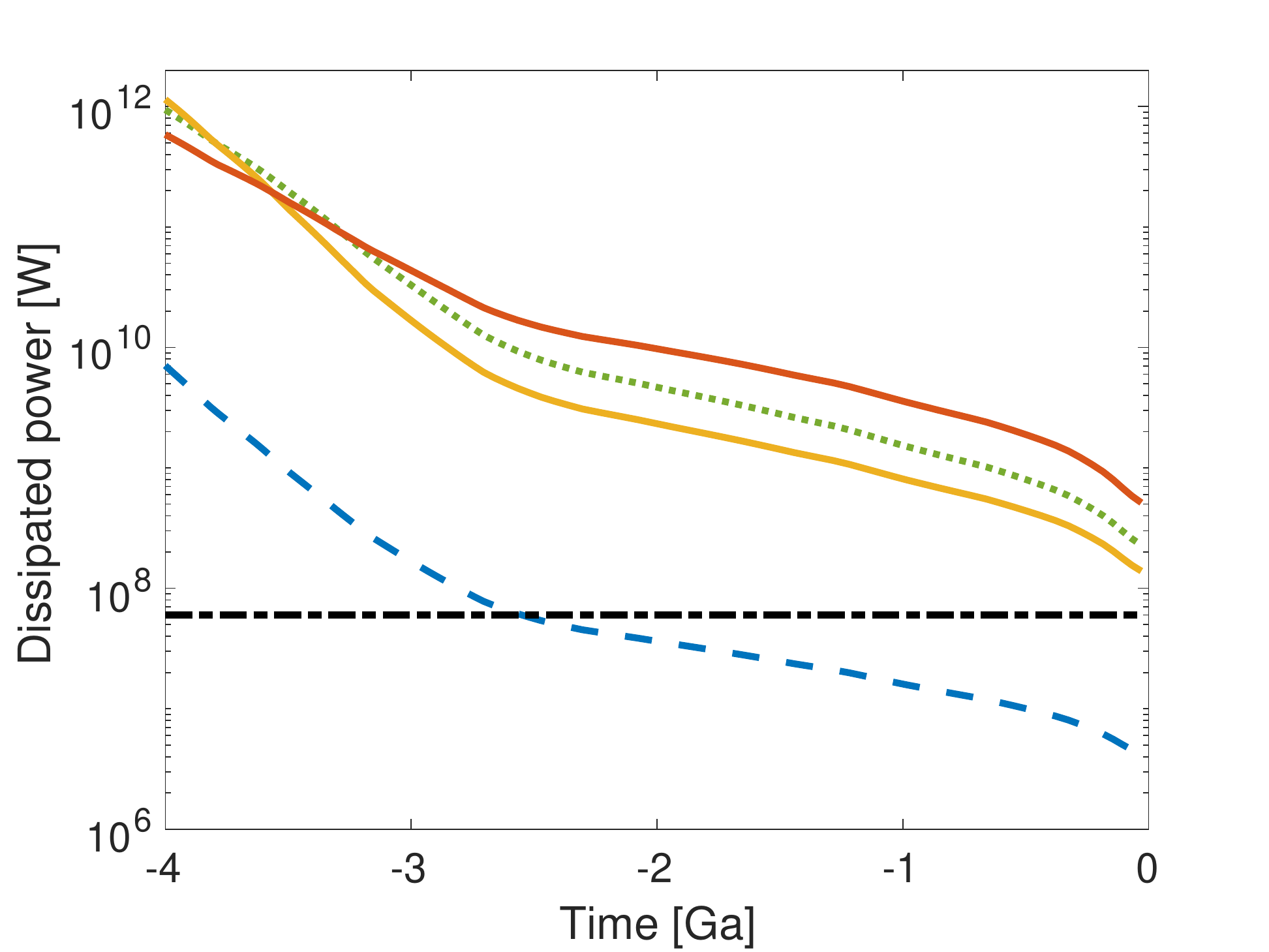}
\caption{Dissipated (dimensional) power in the lunar core given by models, compared to the current dissipation (horizontal dash-dotted line) of $\sim 76$ MW. Laminar model as a dashed line, model of \cite{yoder1981free} as a dotted line, and the model of \cite{sous2013friction} as solid lines (smooth CMB  and $z_r=10^{-5}$ for the lowermost and the uppermost ones at $-1$ Ga).}
\label{fig:MoonDissip2}
\end{center}
\end{figure}

Considering now the lunar core, we take into account a polar flattening of $2.5 \times 10^{-5}$ by solving the torque balance (\ref{eq_axi_P_x})-(\ref{eq_axi_P_z}) with the lunar parameters. The dimensional dissipations are shown in Fig.\ref{fig:MoonDissip2} for the different models of viscous torques discussed above. The lunar fluid dissipation currently observed in the Lunar Laser Ranging (LLR) data is 60 MW according to \cite{williams2001lunar}. Since the dissipation is proportional to the so-called fluid core coupling parameter K/C \cite[see e.g.][]{williams2001lunar}, this value can be updated to 88 MW and 76 MW using the more recent values of K/C respectively given in \cite{williams2014lunar} and \cite{williams2015tides}. Recovering that a laminar dissipation is not consistent with the currently observed dissipation \citep{williams2001lunar}, we also show here that the dissipation is expected to be turbulent during the whole lunar history.  Note that the  turbulent upper bound (\ref{eq:k96}) of \cite{kerswell1996upper} gives much larger dissipation ($> 10^{14}$ W in Fig.\ref{fig:MoonDissip2}), which shows that our turbulent model based on \cite{sous2013friction} is consistent with this theoretical upper bound. Note finally that taking an inner core into account, even with $\eta=0.7$ \citep{weber2011seismic}, does not strongly modify the dissipation (modification by a less than $56\%$ for a smooth CMB).

\section{Conclusion}
\label{conclusion}

In this study, we have characterized the precession driven instabilities necessary to sustain magnetic fields in planetary cores.
At larger viscosities, the boundary layers remain stable, while parametric resonances (CSI) lead to bulk instabilities.
At lower viscosities, the Ekman boundary layer becomes highly turbulent while the CSI is still stirring the bulk.
Varying the inner-core size, we characterize both the evolution of the onset and the dissipation with the shell aspect ratio $\eta$.
We find that it influences only weakly the onset, which is compatible with experimental findings in a librating ellipsoid \citep{lemasquerierlibration}.
While our simulations are still dominated by laminar dissipation, the dissipation in planetary cores may be governed by turbulence in the boundary layers.
Based on our numerical results and the experimental work of \cite{sous2013friction}, we have derived a self-consistent model of the dissipation in precession spherical shells, including both turbulent friction, inner-core and rotation effects.
Extrapolating our model to the lunar core, we predict dissipation compatible with the observed LLR data, with little sensitivity to the size of an inner-core.

Adding the magnetic field, we examine the dynamo action over a wide range of parameters towards the planetary regime ($E,Pm,Po \ll 1$).
At the lowest investigated viscosity $E=10^{-5}$ we found a self-sustained magnetic field at $Pm=0.3$ and $Po=0.007$.
Besides the laminar dynamos at high viscosity which are not relevant for planets, three types of dynamo behaviours emerge at low viscosity ($Pm<1$, $E\leq 10^{-4}$):
(i) stable dynamos, where the magnetic energy reaches a statistically steady state with low to moderate fluctuations;
(ii) intermittent dynamos, where the system oscillates between two states with different mean magnetic energies corresponding to two slightly different directions of the fluid rotation axis;
(iii) self-killing dynamos, where the magnetic field grows exponentially, saturates but finally decays.
At low viscosity, two or three large-scale cyclonic vortices (LSV) are observed during the initial exponential growth of the magnetic field in all three cases.
We suspect that far from their onset, LSV can withstand the Lorentz force leading to stable dynamos.
Furthermore, in the regime of $Po < 0.1$, stable dynamos without LSV have been seen only with $Pm \geq 1$.
For the parameters investigated here, the presence of LSV allows low $Pm$ dynamos, while small-scale turbulent fluctuations are detrimental to dynamo action.
Our results suggest that LSV play a key role in magnetic field generation in (spherical) planetary cores.

Despite the large number of simulations, predictive scaling laws remain elusive for the kinetic energy stored in the instabilities, for the onset of dynamo action, and for the magnetic field strength.
These last two quantities have a non-monotonic behaviour when varying the key control parameters $E$ and $Po$, and an asymptotic regime has yet to be reached.
\cite{nore2014pamir} and \cite{goepfert2018mechanisms} also draw similar conclusions for dynamos in a precessing cylinder and cube, respectively.
This suggests that the non-monotonic behaviour is not linked to the spherical shape, but rather to the precession itself.
For $Pm \leq 2$, the magnetic energy seems capped by the turbulent kinetic energy, in contrast to convective dynamos where magnetic energy overcomes kinetic energy as viscosity is lowered \citep[e.g.][]{schaeffer2017geodynamo}.
Predicting the turbulent fluctuation level is an outstanding issue, as only a few of our simulations reach the regime where departures from laminar dissipation are observed.
Exploring this challenging regime relevant for planetary cores is however a necessary first step towards extracting useful scaling laws -- if such laws exist for this system.
We release our simulation database to enable further investigations and contributions.

None of our dynamos produce a predominantly dipolar field.
Furthermore, at small Ekman number, while the magnetic field is generated in the bulk at the large scale of the LSV, the surface magnetic field is more and more dominated by small scales.
This shift towards small-scale surface fields as the viscosity is lowered was observed for other spherical dynamos driven by the boundaries \citep{monteux2012}.
Indeed, intense small-scale turbulence develops in the boundary layers, shredding the magnetic field to small scales.
This contrasts with convective dynamos for which a wide range of parameters lead to dipolar fields \citep[e.g.][]{kutzner2002stable}.

For a precessing spheroid, a topography driven instability may occur \citep{vidal2017inviscid} while the boundary layers remain stable.
In this case, no small-scale turbulence would shred the magnetic field in the boundary layer, permitting a surface field on the same scale as in the bulk.

\section*{Acknowledgments}

We thank Yufeng Lin for sharing his data with us. We are also grateful to Mathieu Dumberry and two other anonymous reviewers.
Numerical simulations were performed using the XSHELLS code which is freely available at \url{https://bitbucket.org/nschaeff/xshells}.
Our dataset built from our simulations is available at \url{https://doi.org/10.6084/m9.figshare.7017137}.

This work was performed using HPC resource Occigen from CINES under allocations x2016047382, A0020407382 and A0040407382 made by GENCI.
Part of the computations were also performed on the Froggy platform of CIMENT (\url{https://ciment.ujf-grenoble.fr}), supported by the Rh\^ one-Alpes region (CPER07\_13 CIRA), OSUG@2020 LabEx (ANR10 LABX56) and Equip@Meso (ANR10 EQPX-29-01).
NS and DC were supported by the French {\it Agence Nationale de la Recherche} under grant ANR-14-CE33-0012 (MagLune).
ISTerre is part of Labex OSUG@2020 (ANR10 LABX56).
RL would like to acknowledge support from the European Research
Council (ERC Advanced grant 670874 ROTANUT) and the HPC resources of the  Royal Observatory of Belgium.



\appendix

\section{Theoretical solutions for the forced base flow} \label{appendixA}

In this appendix, we consider the usual planetary relevant limit $Po \ll 1$ considered in the literature. In this limit, the unit of time $\Omega_o^{-1}$ corresponds to the usual unit of time $\Omega_s^{-1}$.

\subsection{The model of Busse (1968)} \label{appendixA1}
In the frame of precession, the three Cartesian components $(\Omega_x,\Omega_y,\Omega_z) $ of the dimensionless fluid rotation vector $\boldsymbol{\Omega}$ are governed by the three following theoretical equations (e.g. \cite{Noir2003,cebron2010tilt})
\begin{eqnarray}
    \Omega_z &=& \Omega_x^2+\Omega_y^2 +\Omega_z^2 , \label{eq:NoSpinUp}\\
    - P_z\ \Omega_y &=&  ( \lambda_r\ \Omega_x \Omega_z^{1/4}+\lambda_i\ \Omega_y \Omega_z^{-1/4})\sqrt{E} , \label{eq:Busse2} \\
   P_x\ \Omega_y  &=&   -\lambda_r\ \Omega_z^{1/4}\
   (1-\Omega_z)\ \sqrt{E}  \label{eq:Busse3} ,
\end{eqnarray}
with $P_x=Po \sin \alpha $ and $P_z=Po \cos \alpha $ the two dimensionless components of the precession vector $\vect{\Omega_p}$ along x and z (the y-component is zero).
Equations (\ref{eq:NoSpinUp})-(\ref{eq:Busse3}) are exactly the equations (20)-(22) of \cite{Noir2003}, or equations (21)-(23) of \cite{cebron2010tilt} in the particular case of a sphere (no deformation, and $P_y=0$ in their equations). As shown by \cite{Noir2003}, this system of equations is equivalent to the well-known implicit expression (3.19) of  \cite{busse1968steady}. Equation (\ref{eq:NoSpinUp}) is the so-called no spin-up condition (solvability condition 3.14 of \cite{busse1968steady}, or equation (12) of \cite{Noir2003}) given that it forbids any differential rotation along $\boldsymbol{\Omega}$. Equations (\ref{eq:Busse2})-(\ref{eq:Busse3}) are simply obtained from a torque balance (see \cite{Noir2003,cebron2010tilt} for details). 

In equations (\ref{eq:NoSpinUp})-(\ref{eq:Busse3}), we have noted the spin-over damping factor $\underline{\lambda}=\lambda_r+\textrm{i} \lambda_i$, given by
\begin{eqnarray}\label{lambdaInv}
\underline{\lambda}_{inv}^{sphere}=-\frac{3[19(1-\mathrm{i})+9 \sqrt{3}(1+\mathrm{i})]}{28 \sqrt{2}} \approx -2.62+ 0.258 \textrm{i} 
\end{eqnarray}
for a spherical container ($\eta=0$), in the inviscid limit $E \ll 1$. For finite values of $E$, \cite{noir2001numerical} has obtained empirically $\lambda_r \approx -2.62 -1.36 E^{0.27}$, and a fit of the results of \cite{hollerbach1995oscillatory} gives $\lambda_i \approx 0.258 +1.25 E^{0.21}$. 

Even if equations (\ref{eq:NoSpinUp})-(\ref{eq:Busse3}) are obtained without any inner core, corrections have been proposed for the case of the sphere to take a spherical inner core into account. Using the dimensionless inner radius $r_i$, it has been proposed to simply modify $\underline{\lambda}$ by the factor $(1+\eta^4)/(1-\eta^5)$ for a no-slip inner core \cite{hollerbach1995oscillatory}, and by $1/(1-\eta^5)$ for a stress-free inner core \cite{tilgner2001fluid}. Using the results of \cite{hollerbach1995oscillatory} for the spherical shell, the corrections for viscous and aspect ratio effects can be combined following
\begin{eqnarray}
\underline{\lambda} \approx \left[ \underline{\lambda}_{inv}^{sphere}-1.36 E^{0.27} + \mathrm{i}   \, 1.25 E^{0.21}\right] \frac{1+\eta^4}{1-\eta^5}, \label{eq:lambda}
\end{eqnarray}
with $\underline{\lambda}=\lambda_r+\textrm{i} \lambda_i$.

\subsection{Approximate explicit solution} \label{appendixA2}
 As shown by \cite{Noir2013}, (\ref{eq:NoSpinUp})-(\ref{eq:Busse3}) can be obtained as fixed points of a dynamical model for $\boldsymbol{\Omega}$, given by  (see equations A14-A 16 of \cite{Noir2013} for a spheroid)
\begin{eqnarray}
\frac{\partial \Omega_x}{\partial t}&=&P_z\Omega_{y}-(1-\gamma)\left[P_z\Omega_{y}+\Omega_{y}\Omega_{z}\right]+ \mathcal{L} \boldsymbol{\Gamma}_{\nu}\cdot \unit{x}, \label{eq_axi_P_x}\\
\frac{\partial \Omega_y}{\partial t}&=&P_x\Omega_{z}-P_z\Omega_{x}+(1-\gamma)\left[P_z\Omega_{x}+\Omega_{x}\Omega_{z}\right]+\mathcal{L} \boldsymbol{\Gamma}_{\nu}\cdot \unit{y}, \label{eq_axi_P_y}\\
\frac{\partial \Omega_z}{\partial t}&=&-P_x\Omega_{y}-(1-\gamma) P_x\Omega_{y}+ \mathcal{L} \boldsymbol{\Gamma}_{\nu}\cdot \unit{z}, \label{eq_axi_P_z}
 \end{eqnarray}
where $\gamma=(2a^2)/(a^2+c^2)$ represents the ratio of the polar to equatorial moment of inertia, where $\boldsymbol{\Gamma}_{\nu}$ is the viscous torque, and where $\mathcal{L} $ is a 3x3 matrix given by equation (A3) of \cite{Noir2013}. For the spherical shell, $\mathcal{L} $ reduces to  $\mathcal{L}=15 \delta_{ij} /(8 \pi (1-\eta^5))$, using the Kronecker delta $\delta_{ij}$. As detailed by \cite{cebron2015bistable}, equations (\ref{eq:NoSpinUp})-(\ref{eq:Busse3}) are then recovered with 
\begin{eqnarray}\label{viscousSO}
\mathcal{L} \boldsymbol{\Gamma}_{\nu}=\sqrt{ \Omega E}\left[ 
\lambda_r \left( 
 \begin{array}{ccc}
      \Omega_x \\
      \Omega_y\\  
      \Omega_z-1\\
   \end{array}
   \right)+
 \frac{\lambda_i}{\Omega}\left(   
  \begin{array}{ccc}
     \Omega_y\\
     -\Omega_x\\
      0\\
   \end{array}
   \right)
   \right]. \label{eqNoSUP}
\end{eqnarray}
To obtain tractable analytical solutions, we need a simpler set of equations. We thus follow \cite{Noir2013} who linearize $\mathcal{L} \boldsymbol{\Gamma}_{\nu}$ by assuming that the fluid rotates at the same rate than the boundaries, i.e. $ \Omega=1$,  which gives
\begin{eqnarray}\label{addhocvisc}
\mathcal{L} \boldsymbol{\Gamma}_{\nu}=\sqrt{ E }\left[ 
\lambda_r \left( 
 \begin{array}{ccc}
      \Omega_x \\
      \Omega_y\\
      \Omega_z-1\\
   \end{array}
   \right)+
 \lambda_i \left(   
  \begin{array}{ccc}
     \Omega_y\\
     -\Omega_x\\
      0\\
   \end{array}
   \right)
   \right].
\end{eqnarray}

Focusing on stationary solutions of equations (\ref{eq_axi_P_x})-(\ref{eq_axi_P_z}) in the sphere ($\gamma=1$) with the viscous term (\ref{addhocvisc}), the explicit solution for $\boldsymbol{\Omega}$ is then
\begin{eqnarray}
\Omega_x &=& \frac{[\lambda_i+\chi \cos \alpha] \, \chi \sin \alpha }{\chi (\chi +2 \lambda_i \cos \alpha )+|\underline{\lambda}|^2}, \label{eq:sphereX} \\
\Omega_y &=& -\frac{\chi  \lambda_r \sin (\alpha)}{\chi (\chi +2 \lambda_i \cos \alpha )+|\underline{\lambda}|^2}, \\
\Omega_z &=& \frac{\chi ( \chi \cos^2 \alpha +2 \lambda_i \cos \alpha)+|\underline{\lambda}|^2}{\chi (\chi +2 \lambda_i \cos \alpha )+|\underline{\lambda}|^2}, \label{eq:sphereZ}
\end{eqnarray}
where $\chi=Po/\sqrt{E}$, and $|\underline{\lambda}|^2=\lambda_r^2+\lambda_i^2$. The differential rotation $\epsilon$ of the fluid with the boundary is thus: 
\begin{eqnarray}
\epsilon=  \frac{| \chi \sin \alpha |}{\sqrt{\chi (\chi +2 \lambda_i \cos \alpha )+|\underline{\lambda}|^2}}. \label{eq:RotDif} 
\end{eqnarray}
Note that these explicit expressions for $\boldsymbol{\Omega} $ allows to clarify quantitatively various observations, previously noticed in the literature. For instance, \cite{Noir2013} have considered the resonance of $ \epsilon $ for a fixed Rossby number $Ro=Po \sin \alpha$, i.e. the value of $Po$ where $ \epsilon$ is maximum when $Ro$ is maintained constant. They have noticed that the role of $\lambda_i$ is mainly to shift the resonance peak, that the reduced model (\ref{addhocvisc}) always gives at $Po=0$ for the sphere when  $\lambda_i=0$ is assumed. We can thus use expressions (\ref{eq:sphereX})-\ref{eq:RotDif} to clarify this observation. For a given $Ro$, we obtain that the resonance is indeed reached  for $Po=0$ when $\lambda_i=0$, but the resonance is shifted to
\begin{eqnarray}
\chi=Po/\sqrt{E}=-\lambda_i \sqrt{1+Ro^2/\lambda_i^2} \label{eq:res}
\end{eqnarray}
when $\lambda_i\neq 0$. It is straightforward to show that, at the resonance, $\Omega_x$ is zero whereas $\Omega_y$ and $\Omega_z$ are respectively maximum and minimum.

One can finally notice that the additional hypothesis $\lambda_i=0$, considered e.g. by \cite{Noir2013} and \cite{cebron2015bistable}, allows to simplify equations (\ref{eq:sphereX})-(\ref{eq:sphereZ}) into
\begin{eqnarray}
\boldsymbol{\Omega} &=& \left( \frac{1}{2} \frac{\sin 2 \alpha }{1+x^2}, -\frac{x\sin \alpha}{1+x^2}, \frac{x^2+\cos^2 \alpha}{1+x^2} \right) , \label{eq:sphereZ3}
\end{eqnarray}
which gives
\begin{eqnarray}
\epsilon= \frac{| \sin \alpha |}{\sqrt{1+x^{2}}},  \label{eq:resli0}
\end{eqnarray}
where $x=\lambda_r/\chi$. Note that $\Omega_x$, $\Omega_z$ and $\epsilon$ are maximum for $x=0$, whereas $\Omega_y$ is maximum for $x= \pm 1$. It is interesting to note that equation (\ref{eq:resli0}) is exactly equation (53) of \cite{williams2001lunar}, but we manage here to obtain an explicit expression for $x$.

\end{document}